\documentclass[twocolumn]{aastex631}

\usepackage{amsmath}	

\begin{document}


\title{Kilonova Spectral Inverse Modelling with Simulation-Based Inference: An Amortized Neural Posterior Estimation Analysis}

\correspondingauthor{P.Darc}
\email{phelipedarc@cbpf.br}

\author{P.Darc}
\affiliation{Centro Brasileiro de Pesquisas Físicas \\
Rua Dr. Xavier Sigaud 150, CEP 22290-180, \\
Rio de Janeiro, RJ, Brazil}

\author{C. R. Bom}
\affiliation{Centro Brasileiro de Pesquisas Físicas \\
Rua Dr. Xavier Sigaud 150, CEP 22290-180, \\
Rio de Janeiro, RJ, Brazil}

\author{B. Fraga}
\affiliation{Centro Brasileiro de Pesquisas Físicas \\
Rua Dr. Xavier Sigaud 150, CEP 22290-180, \\
Rio de Janeiro, RJ, Brazil}

\author{C. D. Kilpatrick}
\affiliation{Center for Interdisciplinary Exploration and Research in Astrophysics (CIERA) \\
Northwestern University, Evanston, \\
IL 60201, USA}




\begin{abstract}
Kilonovae represent a category of astrophysical transients, identifiable as the electromagnetic observable counterparts associated with the coalescence events of binary systems comprising neutron stars and neutron star-black hole pairs.
They act as probes for heavy-element nucleosynthesis in astrophysical environments. These studies rely on inference of the physical parameters (e.g., ejecta mass, velocity, composition) that describe kilonovae based on electromagnetic observations. This is a complex inverse problem typically addressed with sampling-based methods such as Markov-chain Monte Carlo (MCMC) or nested sampling algorithms. 
However,  repeated inferences can be computationally expensive due to the sequential nature of these methods. This poses a significant challenge to ensuring the reliability and statistical validity of the posterior approximations and, thus, the inferred kilonova parameters themselves. We present a novel approach: Simulation-Based Inference (SBI) using simulations produced by \texttt{KilonovaNet}. Our method employs an ensemble of Amortized Neural Posterior Estimation (ANPE) with an embedding network to directly predict posterior distributions from simulated spectral energy distributions (SEDs). We take advantage of the quasi-instantaneous inference time of ANPE to demonstrate the reliability of our posterior approximations using diagnostics tools, including coverage diagnostic and posterior predictive checks.  We further test our model with real observations from AT\,2017gfo, the only kilonova with multi-messenger data, demonstrating agreement with previous likelihood-based methods while reducing inference time down to a few seconds. The inference results produced by ANPE appear to be conservative and reliable, paving the way for testable and more efficient kilonova parameter inference.

\end{abstract}


\section{Introduction}\label{sec:introduction}

The coalescence of  binary neutron stars (BNS) and neutron star/black hole-neutron star (NSBH) systems are among the sources of gravitational waves (GW) observed by Advanced LIGO/Virgo \citep{Aasi_2015, Acernese_2015} and can power bright electromagnetic (EM) transients.
The detection of the GW signal GW170817 by the LIGO and Virgo Collaborations is a landmark of the multi-messenger astronomy era \citep{abbott2017search,abbott2020prospects}.
 The GW170817 event encapsulates the scientific advantages of multi-messenger astrophysics; it was accompanied by a cascade of electromagnetic  signals recorded by telescopes in space and all over the world, across the entire EM spectrum \citep{abbott2017search,Kilpa_Shappee_2017}, from gamma-rays to radio emission \citep{2017alex} along with a strong, well-constrained GW signal \citep{abbott2017search}. 
 The so called multi-messenger observations of binary neutron-star (BNS) mergers, which employ different probes to observe the same astrophysical process,  offer insights into the properties of matter under extreme conditions and can be used to infer parameters of the BNS merger and constrain the neutron star equation of state \citep{Radice_2018, Margalit_2017}, as well as the expansion rate of the Universe both from standard sirens or dark sirens perspectives \citep{alfradique24,hubble_kn,darksiren1,2017Natur.551...85A}. In particular, studying the optical and infrared spectra of these EM counterparts can help us understand the process behind the mass ejection mechanism \citep{Metzger_2019}, the physical conditions during the merger and its aftermath. Furthermore, the spectral analysis contributes to studying $r$-process nucleosynthesis and pinpoint the sites where heavy element production occurred over cosmic history \citep[e.g.,][]{Kasen_2017}. 
 
In a neutron star merger, neutron-rich ejecta is released \citep{1974ApJ...192L.145L} and undergoes rapid neutron capture ($r$-process) nucleosynthesis as it decompresses into its environment, enriching our universe with rare heavy elements such as gold and platinum \citep{Metzger_2019}.  
Radioactive decay of these unstable nuclei powers a rapidly evolving, approximately isotropic thermal transient known as a “kilonova”  - an ultraviolet-optical-infrared transient, whose brightness peaks $\sim$ two to three days after a merger \citep{Li_1998}.  

Kilonovae are rare and fast events, with only a few have been observed so far, predominately as counterparts to short gamma-ray bursts at redshifts $z>0.1$ where they are difficult to be observed at optical wavelengths \citep{Bom_2024,Rastinejad21}. Only recently, they were also associated to Long gamma-ray bursts \citep{yang2024lanthanide,Rastinejad22,levan2023}. Therefore, exploring the parameter space of kilonova and their effects on their observable properties is done with simulations, which provide detailed models of ejected matter during mergers \citep{Dietrich_2017modelying, Anand_2020, Kasen_2017}. However, these simulations involve detailed particle physics, general relativity, hydrodynamics, and are therefore computationally complex, taking several hours to produce synthetic observables for one parameter set \citep{bulla2019, Lukosiute_2022}.  These observables then can be compared to real-sky observations data to place constraints on models.


Inferring a kilonova's physical parameters from observational data is a complex inverse problem, which are usually approached using sampling-based inference methods such as Markov-chain Monte Carlo (MCMC) and nested sampling techniques \citep{Cranmer_2020, montel2023scalable}. However, their sequential nature is often an obstacle to rapid, scalable, and testable inference \citep{Cole_2022}.
First,  the time needed to reach convergence scales poorly with the dimensionality of the explored parameter space. This is attributed to the fact that sampling-based inference approaches require sampling the full joint posterior. This latter point is particularly troublesome for problems with large numbers of nuisance parameters, whose posteriors are typically not of direct interest but still require computation, and it arises independently of whether the likelihood is known. 
Second, assessing reliability and statistical rigor in the posterior approximations generated by sampling-based algorithms poses a considerable challenge.  Statistical validation based on repeated inferences, such as simulation-based calibration \citep{talts2020validating} or expected coverage \citep{hermans2022trust} are not feasible in a reasonable time. 
The continuous improvement in computational resources has resulted in a rapid increase in available hydrodynamical models for both supernovae and kilonovae \citep{Dietrich_2020sim}. However, the upcoming Vera C. Rubin Legacy Survey of Space and Time \citep[LSST;][]{verarubin2019ApJ...873..111I} and Nancy Grace Roman infrared space telescope are expected to increase the number of detected BNS events significantly~\citep{Andreoni24}. Consequently, more complex radiative transfer models will emerge, accompanied by new and advanced simulators developed with a broader range of parameters, vastly increasing the need for computational resources for kilonova discovery and characterization. 

 
A new approach capable of dealing with these models' high dimensionality and increasing complexity is necessary. Likelihood-free inference (LFI), or simulation-based inference \citep[SBI;][]{Cranmer_2020} is a rapidly developing class of inference methods that offers alternatives for many applications \citep[see][and references therein]{Cranmer_2020}. The SBI method bypasses the need for an explicit likelihood by training an estimator on simulated data, “learning” to approximate the likelihood or directly estimating the posterior distribution. These models can leverage the use of powerful density estimators such as normalizing flows \citep[NF:][]{papamakarios2021normalizing} and mixture density networks \citep[MDN:][]{papamakarios2018fast}.

In this work, we propose to make use of neural posterior estimation \citep[SNPE-C/NPE;][]{snpe-apt,deistler2022truncated} for rapidly characterizing kilonova spectra.  This approach is based on SBI and uses deep learning to amortize the posterior estimation and infer the physical parameters of simulated kilonovae. The \textit{amortization} means that the trained network has learned not only the marginal posteriors for the observed data but also the marginal posteriors for any data supported by the prior. By adopting this approach, the inference process can be completed within seconds for typical low-resolution optical spectra. Our model is testable due to an amortized approach, enabling posterior diagnostic assessments through repeated inferences in a short time span. Additionally, it is scalable, as it can learn posteriors exclusively for the parameter subset of interest, making it easily adaptable to high-dimensional problems.

This work is organized as follows. Section \ref{sec:kne}, an overview of kilonova ejecta models and the simulator specification are provided.  Section \ref{sec:sbi} details the SBI and NPE approach for approximate inference. Next, in Section~\ref{sec:kntestresults} we present the simulator setup, training procedure, pre-processing steps, and the results obtained from the simulated dataset. This section also evaluates the faithfulness of the posterior estimation.  Finally, in Section~\ref{sec:gwresults} we apply our model to the AT\,2017gfo data.

\section{The astrophysical parameters of Kilonovae}\label{sec:kne}

\begin{figure}
  \includegraphics[width=0.5\textwidth]{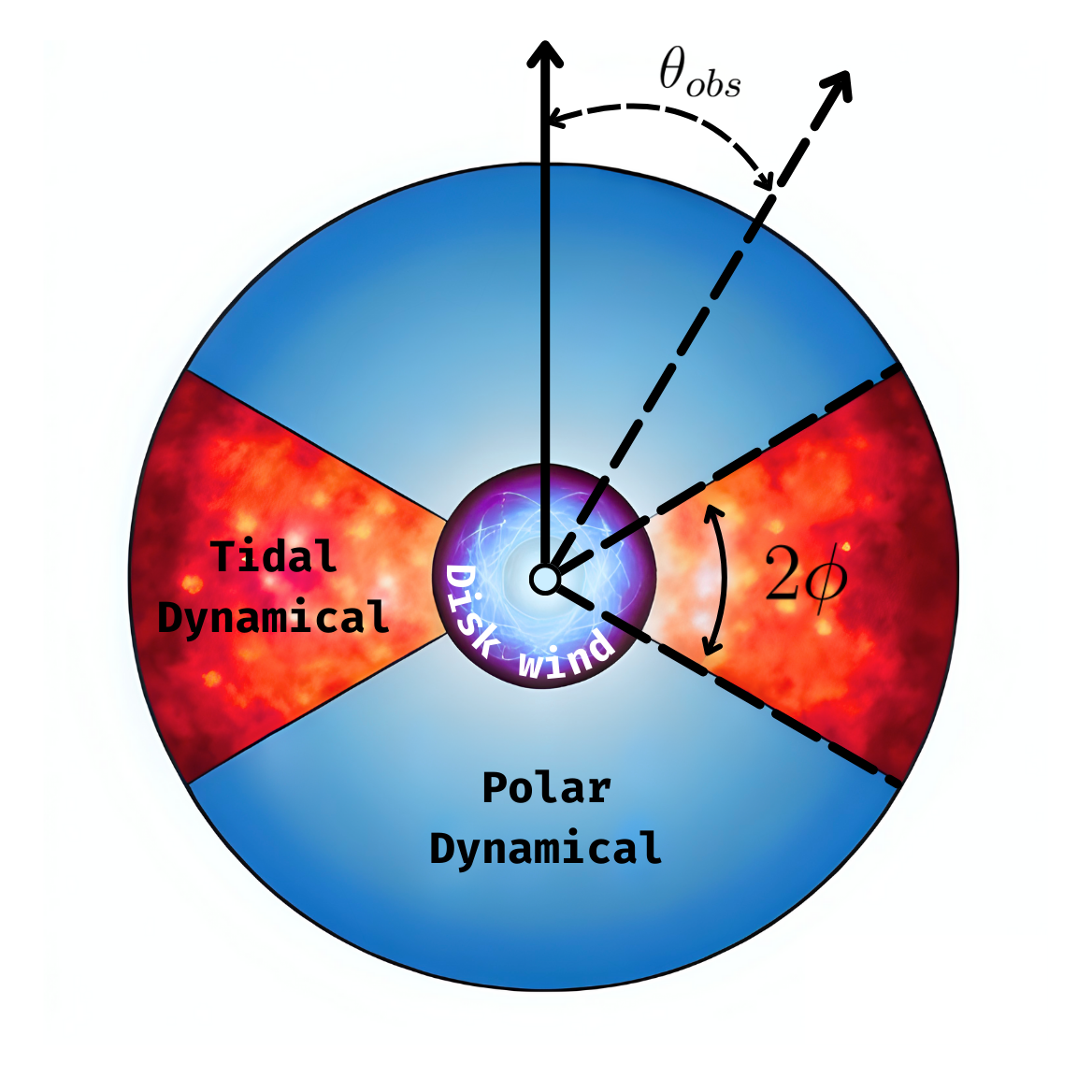}
  \caption{Geometry employed in the kilonova description of Dietrich-based simulation model \citep[see also][]{Dietrich_2020sim}. Different colors refer to the different lanthanide fractions of the individual ejecta components: tidal dynamical (red), polar dynamical (blue), and disk wind (purple).}
  \label{merger}
\end{figure}

Detailed radiative transfer simulations of kilonova spectra play an essential role in multi-messenger astrophysics, providing fundamental information about the elemental abundances, physical conditions, and velocities in kilonova ejecta. Choosing a suitable simulator for parameter inference studies is therefore crucial. Kilonova's spectra and light curves depend strongly on the nuclear yields, neutrino flux, geometric orientations, mass, and velocity of the ejecta \citep{Metzger_2019,Kawaguchi_2020}. State-of-the-art radiative transfer simulations \citep[e.g.,][]{Dietrich_2020sim,Anand_2020} use these parameters to output spectral energy distributions with variable time and spectral resolution. However, when training an amortized NPE model it's necessary to have a large number of simulations that populate the whole parameter space at a resolution comparable to the observed data, making traditional models based on detailed simulations computationally expensive.

To tackle this issue, we employed \texttt{KilonovaNet} \citep{Lukosiute_2022}, a conditional variational autoencoder \citep[cVAE,][]{kingma2022autoencoding} designed to generate surrogate kilonova spectra within tens-of-milliseconds. From these spectral energy distributions and arbitrary filter bandpasses, this simulator also generates light curves in the $ugrizy$ broadband and $H J K$ infrared bands. These methods greatly reduce the time required during parameter inference.

\texttt{KilonovaNet} was trained on three different datasets of simulated kilonova spectra from BNS or NSBH mergers \citep{Dietrich_2020sim, Kasen_2017, Anand_2020}. In this work, we focused on the simulations by \citet{Dietrich_2020sim} as they come directly from BNS merger simulations and are more realistic than the simpler parameterization of ejecta mass, velocity,  elemental abundances and opacity by \citet[][]{Kasen_2017}.  In the models we adopt, the parameter sets consist of the mass of the dynamical ejecta ($M_{\rm ej,dyn}$), the mass of the post-merger ejecta ($M_{\rm ej,pm}$), the half-opening angle of the lanthanide-rich tidal dynamical ejecta  $\phi$, and the cosine of the observer viewing angle $\cos(\theta_{\rm obs})$. 

Figure~\ref{merger} shows different components of the ejecta from BNS mergers.
The dynamical mass ($M_{\rm ej,dyn}$) is the mass ejected during the merger phase and is subdivided into two categories: “tidal” and “polar.” The tidal, dynamical ejecta outflows occur during the final stage of the inspiral in both NSBH and BNS systems when the neutron star is disrupted by the gravitational field of the companion and tend to have a lanthanide-rich composition within an angle $\pm\phi$ about the equatorial plane, resulting in ``red'' kilonova emission that peaks at NIR wavelengths \citep{Dietrich_2020sim,Kasen_2017,Metzger_2019}. Polar dynamical ejecta result from shock heating caused by the direct collision of the neutron stars in BNS systems, which lowers the electron fraction in the ejecta and arrests the production of $r$-process elements.  Thus the polar ejecta have a lanthanide-free composition, resulting in ``blue'' kilonova emission that peaks at optical wavelengths \citep{Metzger_2019}.
A second spherical component represents the ejecta released from the merger remnant and debris disk ($M_{\rm ej,pm}$). The half-opening angle indicates a separation of the ejecta into two components: the matter above and below the tidal ejecta are lanthanide-free. Varying the half-opening angle value has the effect of changing the relative fraction of mass in one compared to the other ejecta component \citep{Dietrich_2020sim}.

\section{Simulation-Based Inference (SBI) of Kilonova Emission}\label{sec:sbi}

The main objective of our Bayesian spectral energy distribution (SED) model is to estimate the posterior distribution $p(\theta \vert x)$ of BNS merger parameters $\theta$, given an observable $x$. For a specific $\theta$ and $x$, we typically evaluate the posterior using Bayes’s rule, $p(\theta \vert x) \propto p(\theta) \times p(x \vert \theta)$ , where $p(\theta)$ denotes the prior distribution and $p(x \vert \theta)$ the likelihood, it denotes the probability that a given set of parameters $\theta$ could have generated the observable $x$.
Traditional methods used in astrophysical parameter inference often rely on approximate likelihoods for posterior estimation. These approximations inherently assume parameter independence and a Gaussian distribution. However, this simplification may not capture the complexity of the problem. Furthermore, the statistical validation of the approximations produced by stochastic algorithms (e.g., Monte Carlo methods) are difficult to assess, mostly because of the time needed to compute a posterior distribution for one observable.

To address these limitations, SBI offers an alternative approach that requires no assumptions about the form of the likelihood.  Rather than relying on the likelihood function to perform inference, simulation-based approaches utilize deep neural networks to parameterize universal density estimators and estimate the posterior distribution.   
Specifically, we employ the neural posterior estimation (NPE) method   (\cite{snpe-apt,papamakarios2018fast,deistler2022truncated}). This method involves training a neural density estimator $p_\phi(\theta \vert x)$ with the parameters $\phi$ and directly estimate the posterior distribution  $p(\theta \vert x)$. The optimization problem is to minimize the expected Kullback-Leiber (KL) divergence between the two distributions for all the observations, that is:

\begin{equation}
 \begin{aligned}
    & \arg \min _\phi \mathbb{E}_{p(x)}\left[\operatorname{KL}\left(p(\theta \mid x)        \| p_\phi(\theta \mid x)\right)\right] \\
    = & \arg \min _\phi \mathbb{E}_{p(x)} \mathbb{E}_{p(\theta \mid x)}\left[\log           \frac{p(\theta \mid x)}{p_\phi(\theta \mid x)}\right] \\
        = & \arg \min _\phi \mathbb{E}_{p(\theta, x)}\left[-\log p_\phi(\theta \mid            x)\right]
\end{aligned}
\end{equation}

The chosen neural density estimator is a mixture density network (MDN; \cite{papamakarios2018fast}). Thus $p_\phi(\theta \vert x)$ takes the form of a mixture of gaussian components, such architecture is capable of representing any conditional distribution arbitrarily accurate — given that the number of components and number of hidden units in the neural network are sufficiently large — while remaining trainable through back propagation. 

\subsection{Amortized Neural Posterior Estimation}

The NPE method offers another critical advantage over others SBI methods: \textit{amortization}.  The core strategy involves initially training a model (training phase)  – specifically, a \textit{density estimator} –  that is  not focused on any particular observation. Instead, it aims to be a versatile estimator that attempts to approximate all posteriors supported by the prior. Once trained, the density estimator can, in the next phase (inference), quickly and continually infer parameters of BNS kilonovae from their spectra. To train the network, we can simulate observables using prior-draw parameters to build a dataset, of kilonova parameters ($\theta_j$) and their respective spectra ($X_j$), and minimize the Loss function over the weights.  Once the density estimator has been trained on simulated data, it can then be applied to empirical data $X_0$ (AT\,2017gfo) to compute the posterior. 

\begin{figure*}
    \centering
    \includegraphics[width=0.9\linewidth]{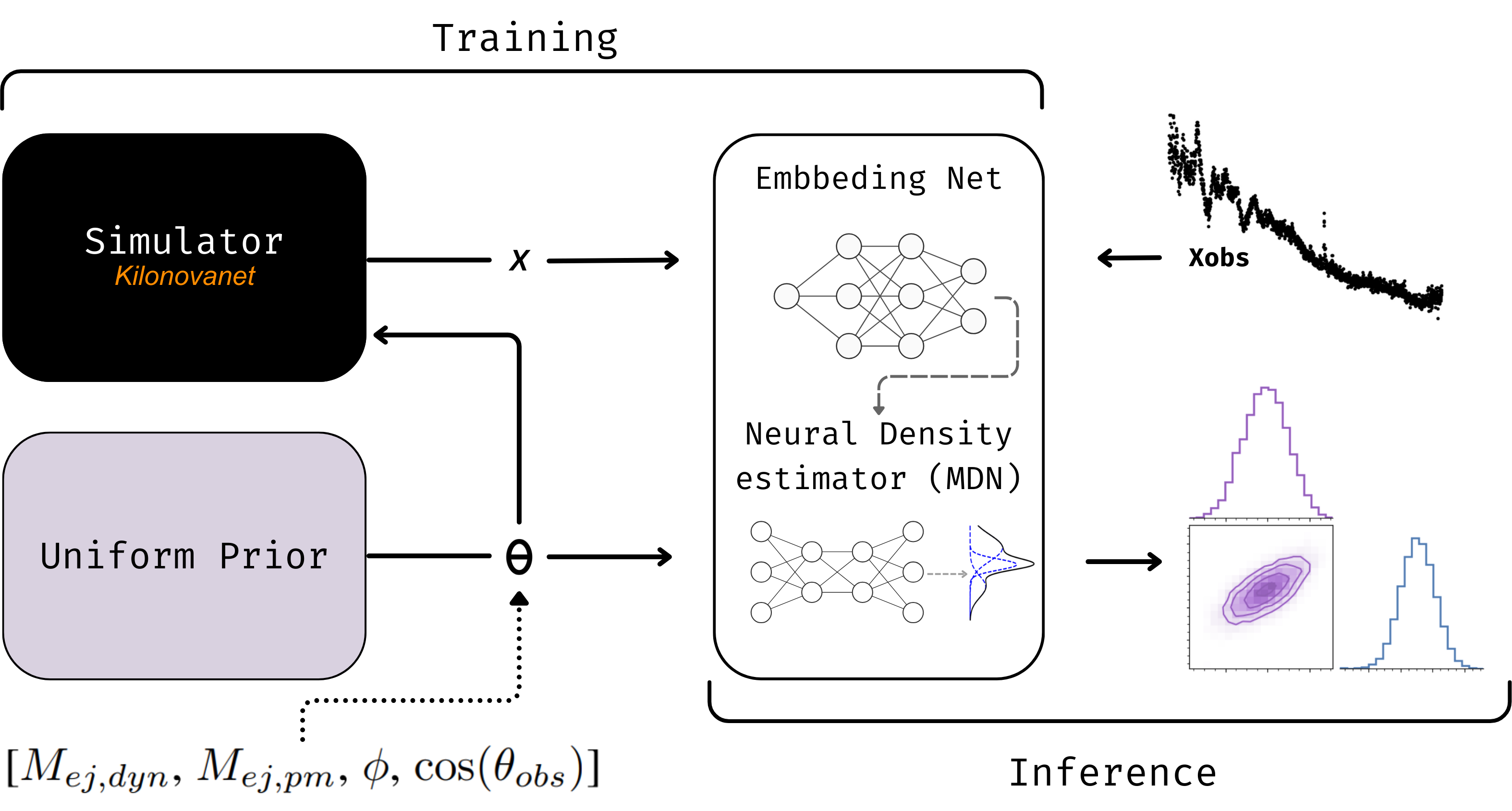}
    \caption{Inference pipeline using Amortized Neural Posterior Estimation — amortized inference enables rapid inference for every new spectrum observed. In the training phase, a neural-network-based density estimator (MDN) learns the probabilistic relationship between the model’s parameters and simulated spectra, at high computational cost. Then, in the inference phase, the trained density estimator takes the observed real data as input and infers the posterior distribution of the parameters, at low computational cost: less than a second.}
    \label{fig:ANPE}
\end{figure*}

Figure~\ref{fig:ANPE} illustrates an overview of the ANPE method using mixture density network (MDN) as the density estimator and \texttt{KilonovaNet} as the simulator to construct posterior distributions for the astrophysical parameters.

\section{Results}\label{sec:kntestresults}

 We start our empirical evaluation of our NPE-based kilonova parameter estimator model by describing the creation of the training data in Section \ref{traininsetup},  accompanied by a description of the neural posterior estimator configuration. 
 
 Section \ref{kntest} displays a demonstration of parameter estimation and spectra retrieval employing NPE. 
Subsequently,  we assess the statistical robustness and precision of our model on new generated simulated test set through a variety of statistical and diagnostic tools. To evaluate the precision and accuracy of our framework, we compare the recovered SED parameter values with the true values and analyze the residual distribution for each parameter. Additionally, to check the health of the calculated posteriors, we also perform posterior predictive checks \citep[PPCs;][]{Gabry_2019}, Coverage test \citep{hermans2022trust} and simulation based calibration \citet[SBC;][]{talts2020validating}.

\vspace{2cm}

\subsection{Training Setup\label{traininsetup}}

\texttt{KilonovaNet }takes a set of parameters [$M_{ej,dyn}$, $M_{ej,pm}$, $\phi$, $\cos(\theta_{obs})$] and a list of days after the merger to produce a spectrum. 
Initial tests showed that parameter estimation using single-day spectra is too broad (i.e., it lacks predictive power and simply recovers our prior) or converges on an incorrect parameter set. 
This phenomenon may arise from the limitation inherent in utilizing a single day,  as the evolving characteristics of a kilonova spectra, such as the overall shape and color evolution, are not intrinsically presented to our model. Consequently, critical information regarding the amount of material ejected and the half-opening angle becomes inaccessible, leading to a parameter degeneracy and subsequently yielding poor constraints. 

Therefore, we combined spectra from three different times after the merger (1.5, 2.5, and 3.5 days) to serve as input. Furthermore, an ensemble of models was trained to encompass various time intervals, specifically [$1.0+x, 2.0+x, 3.0+x$], where $x$ ranged from 0.00 to 0.99, in increments of 0.2. This approach was adopted to address potential variations in observation times given the rapid evolution of kilonova spectra, in particular during the first few days post-merger \citep{Kilpa_Shappee_2017}. 
However, our main focus in this work is to test and validate our method on the AT\,2017gfo event. Therefore, our analysis will primarily center on the model trained for three distinct post-merger time points: 1.5, 2.5, and 3.5 days.

We generate $150,000$ spectra for each day ($450,000$ spectra in total) and cut them to wavelengths between $5000$ and $8000$~\AA. Each spectrum was interpolated to have 550 points. Since the wavelengths for all spectra are the same after interpolating, we only use the flux value as input. To avoid any intrinsic biases that could arise from the simulator, we applied Gaussian smoothing to reduce the noise from the simulator. To map our training set to observations, we added stochasticity to the training set by adding a Gaussian noise corresponding to 10\% of the flux at each point. 
Preliminary tests showed that there is no performance gain by normalizing the flux to zero mean and unit variance, or normalize it between 0 and 1, therefore we did not normalize the spectra's flux. 

We choose uniform priors for all the parameters in the simulations, with the minimum and maximum corresponding to the smaller and larger value coming from the original simulated data \citep{Dietrich_2020sim}. We use the PyTorch-based SBI library \citep{SBItejero-cantero2020sbi}\footnote{https://www.mackelab.org/sbi/}, and their implementation of Amortized Neural posterior estimation (ANPE), with a mixture density network as the density estimator. 


Due to the complexity of these data, we implemented an embedding network in our model. The following baseline architecture for our embedding network was decided:  Three convolutional layers followed by a max pooling layer, two Long Short-Term memory (LSTM) layers and four dense layers. The embedding Network was used to compress the spectrum of shape $(550,3)$ into a vector of $100$ features, which was then used to condition the MDN with respect to the input data. The embedding network's weights were optimized during training together with the density estimator.
The use of Convolutional layers and LSTM layers are essential to extract specific patterns and trends in the spectral data, which can be indicative of certain physical properties, such as the lanthanide composition of the ejecta and the viewing angle \citep{Metzger_2019}.

\begin{deluxetable}{ccc}
\tablewidth{0pc}
\tabletypesize{\footnotesize}
\tablecaption{Embedding Net Architecture with Input Size (550,3)}
\tablehead{\colhead{\textbf{Layer}} &
\colhead{\textbf{Operation}} &
\colhead{\textbf{Output Size}}
}
\startdata 
Conv Block & Conv1d (3 $ \rightarrow$ 128, 2x2, ReLU) & (550, 128) \\ \hline
           & MaxPool1d (2, stride=2) & (275, 128) \\ \hline
Conv Block & Conv1d (128 $ \rightarrow$ 128, 2x2, ReLU) & (275, 128) \\ \hline
           & MaxPool1d (2, stride=2) & (137, 128) \\ \hline
Conv Block & Conv1d (128 $ \rightarrow$ 64, 2x2, ReLU) & (137, 64) \\ \hline
           & MaxPool1d (3, stride=3) & (45, 64) \\ \hline
LSTM & LSTM (64 $ \rightarrow$ 300, bidir)& (45, 600) \\ \hline
LSTM & LSTM (600 $ \rightarrow$ 300, bidir) & (45, 600) \\ \hline
Flatten & Flatten & 2700\\ \hline
FC Block & Linear (2700 $ \rightarrow$ 256, ReLU)& 256 \\ \hline
         & Linear (256 $ \rightarrow$ 128, ReLU) & 128 \\ \hline
         & Linear (128 $ \rightarrow$ 64, ReLU) & 64 \\ \hline
         & Linear (64 $ \rightarrow$ 100) & 100 \\ \hline
Activation & ReLU & - \\ \hline
\enddata 
\end{deluxetable}

Current Bayesian simulation-based inference algorithms, such as NPE, can produce computationally unfaithful posterior approximations \citep[see][]{lueckmann2021benchmarking, hermans2022trust}, thus yielding overconfident posterior approximations, which makes them unreliable for scientific uses cases. 
Following \cite{hermans2022trust} suggestion, We adopted an ensemble approach; instead of training a single model, we train an ensemble composed of five models using a K-Fold technique. The training set was divided into five folds where each fold has the opportunity to be used as a validation set. This allows for a more conservative posterior estimation, the quality of the posterior will be discussed in more details in Section~\ref{coverage}.

Note that for a large number of simulated spectra, we do not expect any performance gain or robustness by using a K-fold, or bootstrapping, or training on five different sets generated by our simulator. We chose the K-Fold approach because it is the most common technique for training ML models, therefore it's well described and validate in many fields of science.

Each model was trained until the loss stopped improving for 60 epochs and with a batch size of 1024. The optimization was carried out through a variant of stochastic gradient descent, namely Adam \citep{kingma2017adam} with a learning rate of 0.0001. 
The model was trained in a Multi-GPU server with 8 RTX 3090 with 24 GB of GPU memory each, with liquid cooling system and 4TB of ram memory. The training process took approximately 3 hours.



\subsection{Retrieving Kilonova Parameters\label{kntest}}

\begin{figure}
     \centering
     \includegraphics[width=1\linewidth]{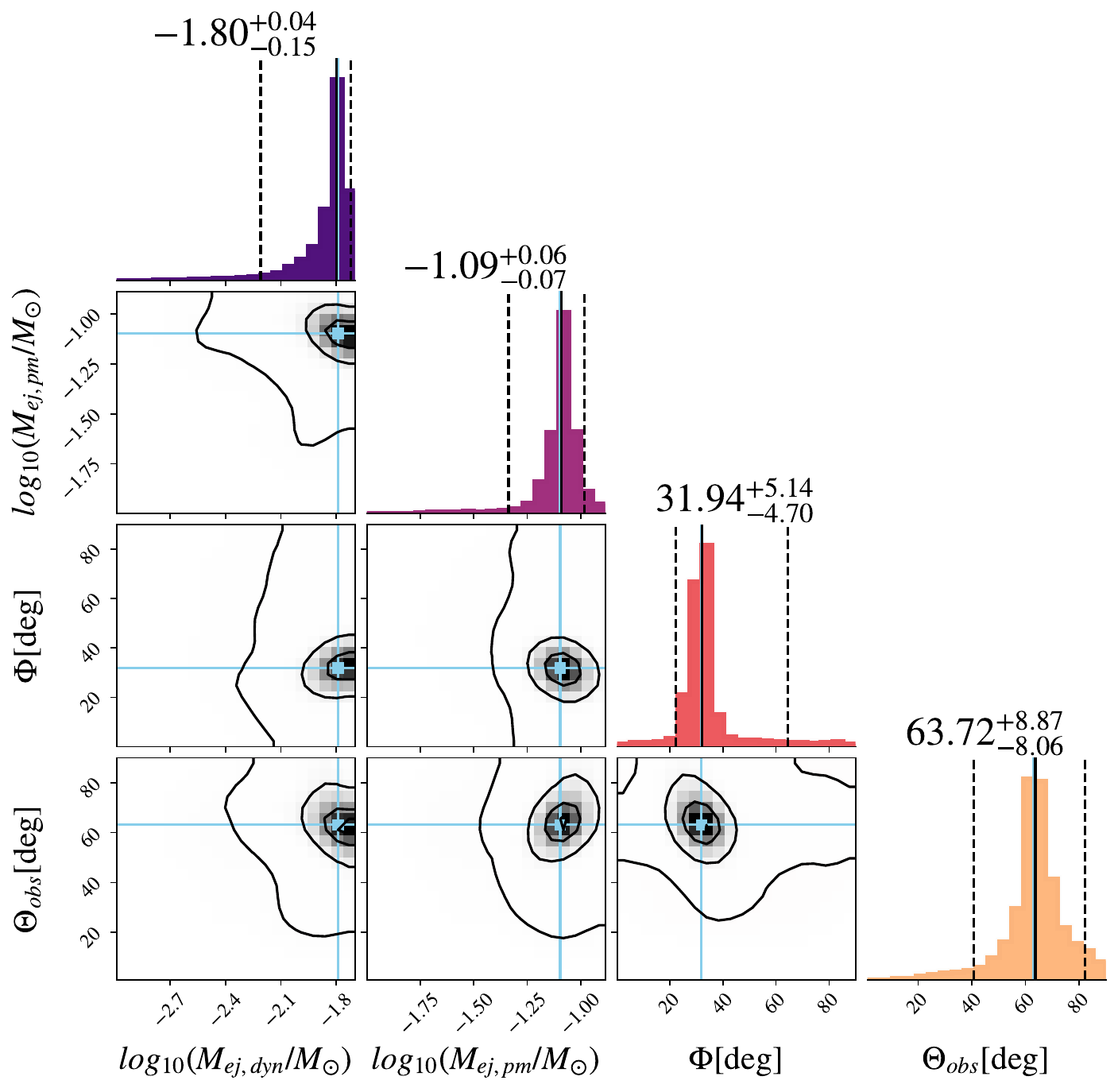}
    \caption{Corner plot for an example spectra ($X_{sim}$) in our test set, it shows the Inferred posteriors for kilonova parameters at 10\%, 32\%, 68\% and 95\% confidence intervals. The median and $90\%$ confidence interval are shown in vertical solid and dashed lines, respectively, and reported above each column. . We observe that the nominal parameters $\theta_{sim}$ (light blue) are accurately retrieved. }
     \label{fig:cornertest}
\end{figure}

We start the evaluation of our NPE model by demonstrating the ability of your model to reconstruct the spectra and perform an accurate and precise retrieval of the kilonova parameters for one sample randomly selected from the test set.  The inference results for the $X_{\rm sim}$ generated using the true parameters $\theta_{\rm sim}$ draw from the test set are summarized in Figure \ref{fig:cornertest}. The corner plot shows 1d and 2d marginal posterior distributions obtained for the benchmark spectrum $X_{\rm sim}$.  Where the contours are the  the 10\%, 32\%, 68\% and 95\% confidence intervals. In the one-dimensional posterior panels, the corresponding 90\% credible intervals (CIs) are given as dashed, black lines, while medians are illustrated as solid lines. We observe that the ground truths, $\theta_{\rm sim}$, are contained within the 1-$\sigma$ credible regions, which we interpret as evidence that our NPE approach is capable of producing reasonable accurate posterior distribution. 
\begin{figure*}
     \centering
     \includegraphics[width=1\linewidth]{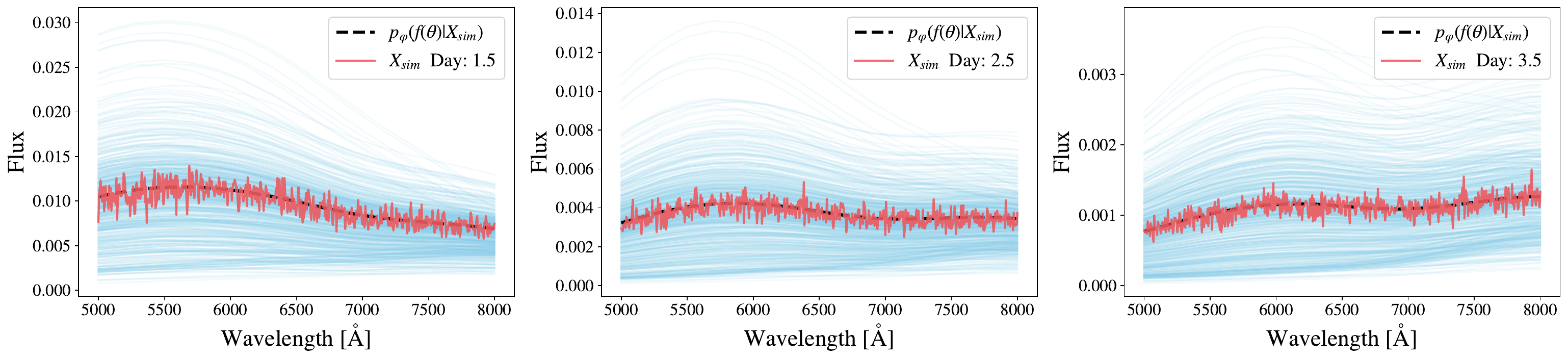}
     \caption{Posterior Predict Check  showcases the comparison between simulated spectra ($X_{sim}$, dark orange line), posterior-predicted spectra ($X_{pp}$, light blue lines), and the spectra reconstructed using the most probable parameters (black dashed line).    }
     \label{fig:ppc}
\end{figure*}

Posterior Predict Check \citep[PPC;][]{Gabry_2019} is a visual tool to check the faithfulness of our inferred posterior. This is accomplished by comparing the input spectra ($X_{\rm sim}$) and the spectra generated ($X_{\rm pp}$), utilizing \texttt{KilonovaNet}, by sampling 600 times the inferred posterior distributions of  $X_{\rm sim}$. The main idea is that if the inference is correct, the spectra generated must be similar to the $X_{\rm sim}$.  Figure \ref{fig:ppc} displays the $X_{\rm sim}$ (Dark orange line), $X_{\rm pp}$ (light blue lines) and the spectra reconstructed using the most probable parameters (black dashed line). PPC is not a validation metric; rather, it function as diagnostic tool for inference. It offer insights into potential biases introduced during the inference.  We can notice that the black dashed line follows  $X_{\rm sim}$ for all the three days after the merger, and the majority of the sampled spectra are very close to the $X_{\rm sim}$.

\begin{figure}
    \centering
    \includegraphics[width=1\linewidth]{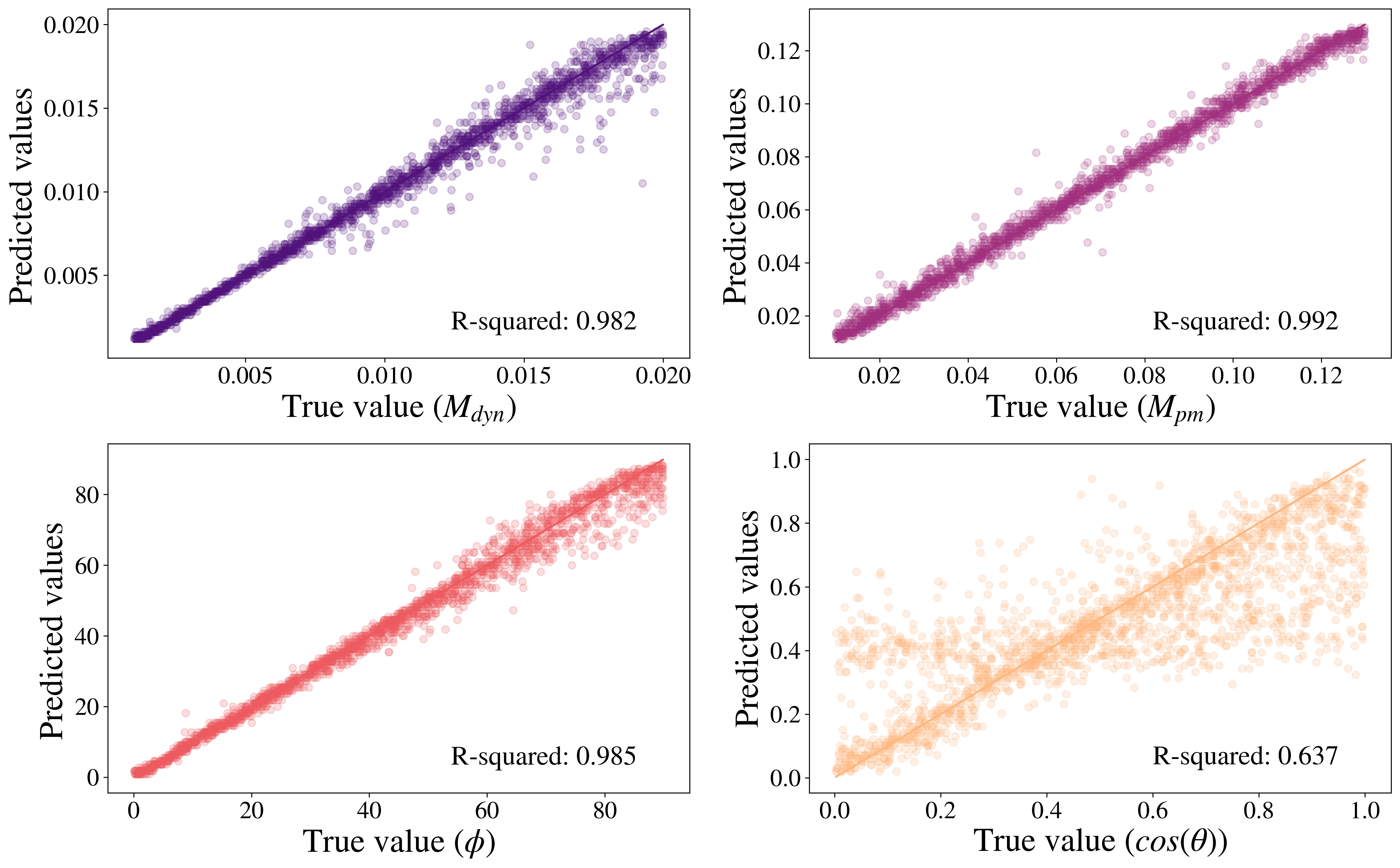}
    \caption{ True vs. recovered values for kilonovas parameters in the test set of 2000 spectra. The diagonal solid line is the ideal scenario where the $\theta_{pred}$ is equal to $\theta_{true}$. This demonstrates the accuracy of the ANPE model across the entire range of priors.}
    \label{fig:predtrue}
\end{figure}

We also applied our NPE method to infer the posteriors distributions for each observation and computed the corresponding most probables values ($\theta_{\rm pred}$) for every observation within the test set ($X_{\rm test}$;$\theta_{\rm true}$). Figure \ref{fig:predtrue}  shows a comparison between the $\theta_{pred}$ and the ground truth values $\theta_{\rm true}$, we achieved a coefficient of determination ($R^2$) for the parameters $M_{\rm ej,dyn}$, $M_{\rm ej,pm}$, $\phi$, $\cos(\theta_{\rm obs})$ of $0.982$, $0.992$, $0.985$, and $0.637$, respectively.  Furthermore, we computed the difference between the  $\theta_{\rm pred}$ and $\theta_{\rm true}$. For a non-biased accurate model, we expect the residual distribution to be centered around zero, with no clear preference for positive or negative values. 

\begin{figure}
    \centering
    \includegraphics[width=1\linewidth]{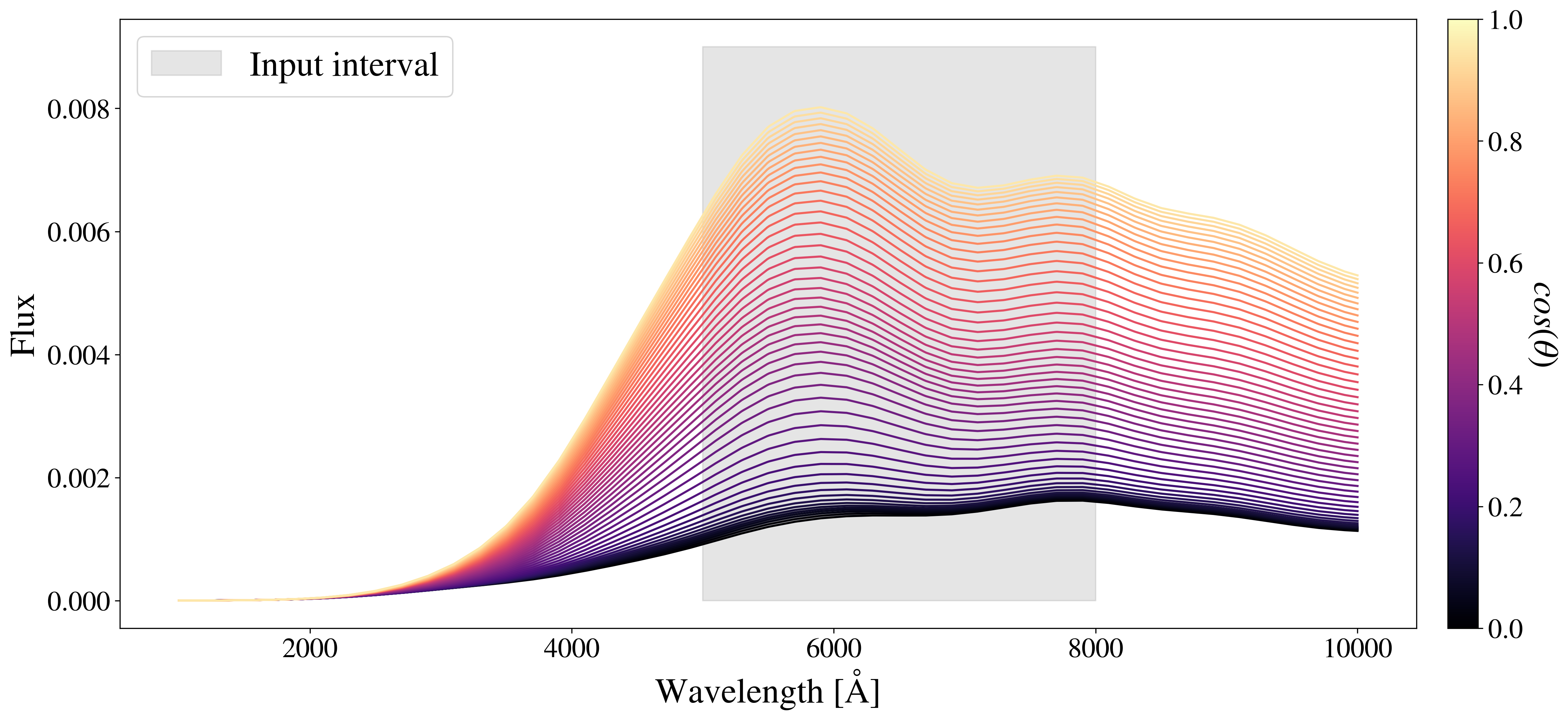}
    \caption{Spectral energy distributions (SEDs) 2.5 days after the merger for a fixed $M_{\rm ej,dyn} =0.0163~M_{\odot}$, $M_{\rm ej,pm} = 0.0803~M_{\odot}$ and $\phi =31.18^{\circ}$ (same value as  $\theta_{\rm sim}$). The SEDs are shown for 50 different viewing angles from the equator (black, edge-on, $\theta_{\rm obs} = 90^{\circ}$) to pole (Light Orange, face-on, $\theta_{\rm obs} = 0^{\circ}$). It is observed that the overall shape exhibit minimal changes, particularly for values of $\theta_{\rm obs}$ near the equator. }
    \label{fig:theta}
\end{figure}

\begin{figure}
    \centering
    \includegraphics[width=1\linewidth]{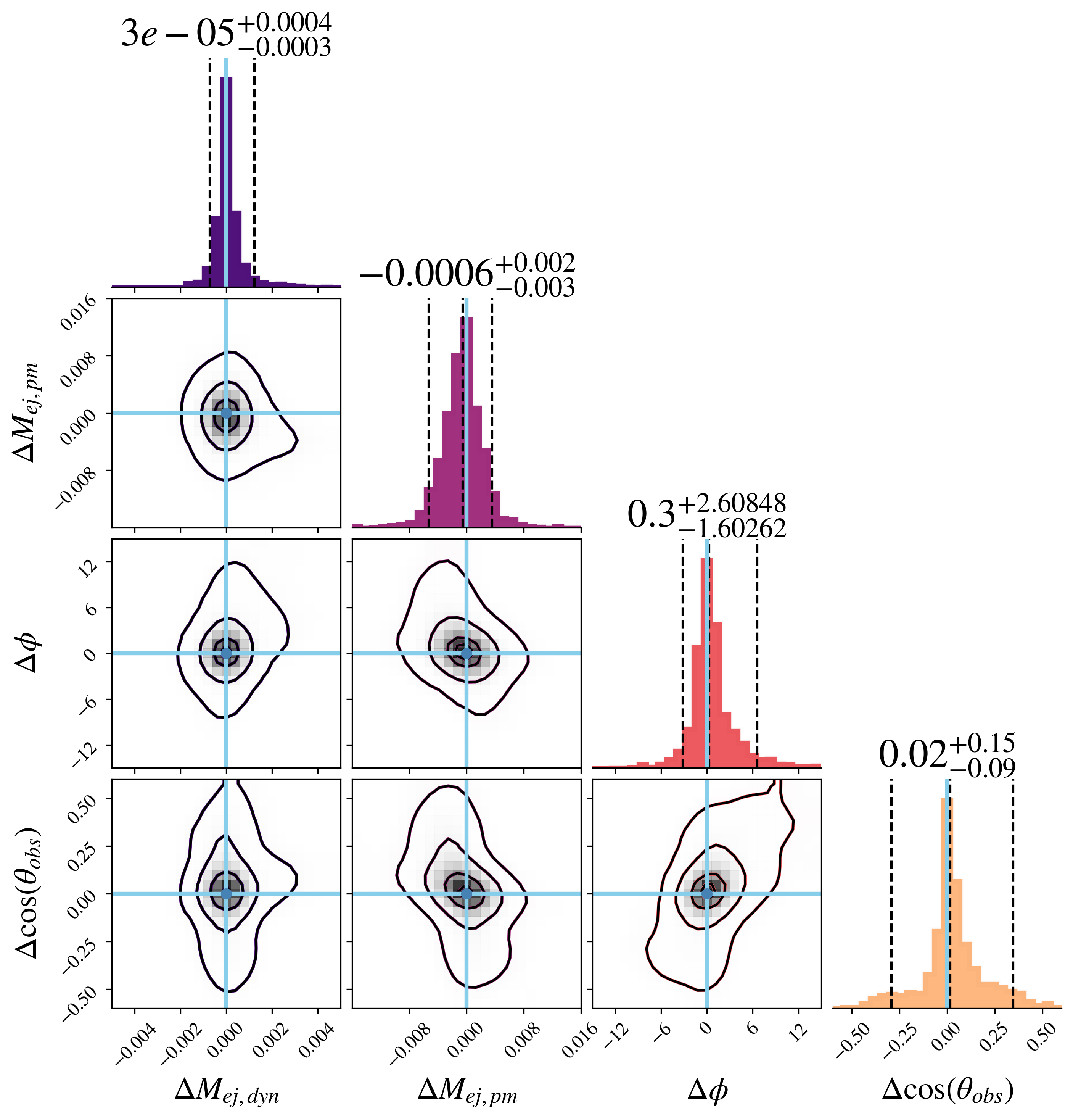}
    \caption{Residual Corner Plot shows the 1d and 2d residual distributions obtained by subtracting $\theta_{\rm true}$ from $\theta_{\rm pred}$. We also show the zero in light blue and the median and $90\%$ interval are shown in vertical dashed lines, and reported above each column. We observe that the residual distribution is approximately centered around zero for all the parameters.  }
    \label{fig:residualplot}
\end{figure}

Figure \ref{fig:residualplot}  displays the contour plot of the residual distribution for each parameter. The marginal distributions show the median value and the $90\%$ interval as dashed lines, the light blue line is a reference which is centered in zero.  For all the parameters, the value is very close to zero, highlighting the absence of a systematic overestimation or underestimation of the inferred parameter. The cosine of the observing angle has a slight underestimation, but the bulk of the residual distributions (orange) is centered around zero.  Thus, we demonstrate accurate and precise parameter recovery across the majority of the simulated range. The constraints on the cosine of the observing viewing angle appear to be broader and less precise compared to the other parameters, particularly for lower values.  This comes from the fact that the overall spectral slope doesn't change significantly as you vary the viewing angle, especially for low values, so it's hard to constrain it from the spectrum. To illustrate this observation, in Figure \ref{fig:theta} we fixed the parameters $M_{\rm ej,dyn}$, $M_{\rm ej,pm}$, $\phi$, and simulate spectra 2.5 days after the merger varying the $\cos(\theta_{\rm obs})$ from 0 to 1, the shaded area is the wavelength interval of our input data. This effect is improved when using a broader wavelength range and more days after the merger, which in most cases are not available in a real scenario, this same effect is evident at 1.5 and 3.5 days after the merger.


\subsection{ Posterior Diagnostic}\label{coverage}

\begin{figure}
    \centering
    \includegraphics[width=1\linewidth]{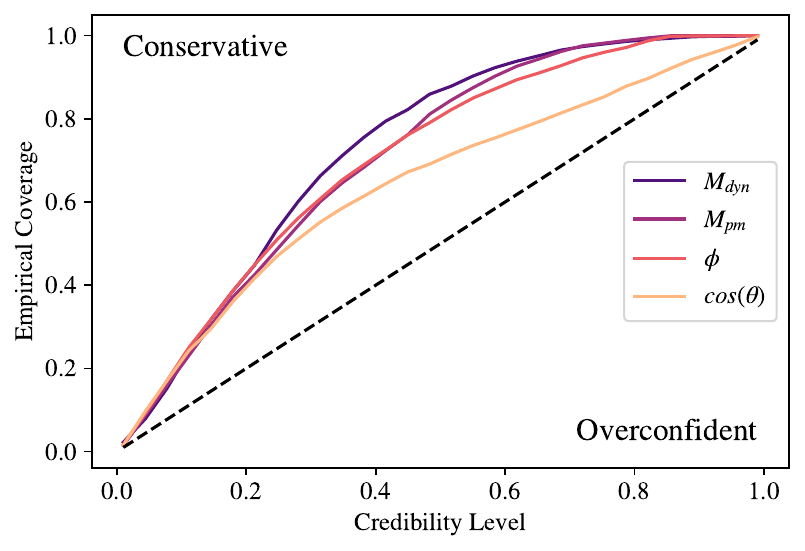}
    \caption{Coverage plot for each kilonova parameter. A conservative posterior estimator will produce curves above the dashed line, and an overconfident posterior estimator will produce curves below the diagonal.}
    \label{fig:coverage}
\end{figure}

An important step to validate our posterior estimates is to measure the quality of the credible regions produced by our Amortized NPE model. The coverage test, as documented by \cite{hermans2022trust}, serves as a diagnostic tool to evaluate whether the uncertainties of the posterior are balanced, i.e., neither over-confident nor under-confident. 

The coverage test initiates by defining  a credible region for the estimated distribution using Highest Posterior Density Regions. This region represents the smallest area that contains at least $100(1 - \alpha)\%$ of the mass of the inferred posterior distribution, establishing an interval corresponding to a specified credibility level ($1-\alpha$). The expected coverage is the frequency with which the true parameter ($\theta_{true}$) value falls within this highest density region, essentially indicating how often it falls inside the calculated interval.

If our model generates well-calibrated distributions, on average, the true value is expected to be enclosed within the computed interval derived from the $1-\alpha$ highest posterior density regions of the posterior distribution $p_{\phi}(\theta \vert x)$ exactly $1-\alpha\%$ of the time. An expected coverage smaller than the credibility level 
$1-\alpha$ indicates that the $1-\alpha$ highest posterior density regions are narrower than anticipated, indicating overconfidence and often unreliable posterior approximations. Conversely, an expected coverage larger than the credibility level $1-\alpha$
implies that the highest posterior density regions are broader than anticipated. In such instances, the posterior approximations are said to be conservative.

The graphical representation of this expected coverage is commonly referred to as coverage plot (see Figure \ref{fig:coverage}). We argue that simulation-based inference should rather produce conservative posterior distributions to guarantee a reliable and meaningful inference. An overconfident distribution would wrongly exclude  probables values for the kilonova parameters, and it could lead to wrong conclusions about the physics behind the merger. While a conservative model would only fail at excluding actually implausible parameter values, resulting only in a loss of statistical power. 
 Figure \ref{fig:coverage} summarizes the expected coverage of $p_\phi(\theta \vert x)$ for credibility levels from 0 to 1 for each kilonova model's parameters.
Our deep ensemble ANPE produces a curve above the diagonal, which provide additional support that our model produces conservative, reliable posteriors.


To extend our posterior diagnostic, we also evaluate the Simulation-Based Calibration (SBC; \cite{talts2020validating}), which is a general procedure for validating inferences from Bayesian algorithms capable of generating posterior samples. In SBC, we draw samples $\theta '$  from the Data averaged posterior (DAP), i.e, the posterior obtained by running inference for our test set. When the posterior approximation is exact,  $\theta '$ is expected to match the prior distribution, i.e, a uniform distribution.  To compare the two distributions, we used the Classifier two sample test (C2TS;\cite{lopezpaz2018revisiting}), In the context of SBI, C2ST has, e.g., been used in \cite{Dalmasso_2020,lueckmann2021benchmarking}.
 This is a nonparametric two-sample test based on training a classifier to differentiated one of the distributions (DAP versus samples from a uniform distribution) by being trained on the other. If one sees values around $0.5$, the classifier was unable to differentiate both distributions, i.e., DAPs are very uniform. If the values are high towards \verb|1|, this matches the case where DAP is very unlike a uniform distribution. 
We report the mean values of [$0.505, 0.491,0.501,0.501$] for the deep ensemble NPE models.  Therefore, we can conclude that our model does not exhibit a strong systematic shifted mean or overconfidence.

The primary benefit of employing amortized NPE lies in the ability to evaluate a substantial number of tests and diagnostics. In a traditional MCMC approach, many of these tests would be impractical and time-consuming. 


\section{Characterizing Spectra from the GW170817 event}\label{sec:gwresults}
On 2017 August 17, AT\,2017gfo was discovered as the optical counterpart to a Gravitational wave event (GW170817; \cite{abbott2017search,Soaressantos2017ApJ...848L..16S}) caused by a binary neutron star merger,  the only optical counterpart to a BNS GW event discovered to date. AT\,2017gfo was localized to the S0 galaxy NGC\,4993 at a distance of 40~Mpc approximately 11~hours after merger \citep{Coulter17}, enabling the first spectra to be obtained within a day of merger. 
 We use spectral data collated in \cite{Kilpa_Shappee_2017}, taken at 1.46, 2.49 and 3.46~days after the merger and obtained using the Low Dispersion Survey Spectrograph-3 (LDSS-3) at the Las Campanas Observatory, Chile (for more details see \cite{Kilpa_Shappee_2017}).
These spectra were pre-processed by passing a uniform filter of size 10 spaxels (corresponding to a step of $54.45$ \AA) in order to be visually similar to the simulated data. 
 
We start this section by employing our ANPE model to retrieve AT\,2017gfo's parameters. We employ the computational tool \texttt{KilonovaNet} to simulate spectra using  the inferred posterior distributions. These simulated spectra are visually compared to both the observed kilonova spectra and spectra simulated using results from previously established likelihood-based methods. Additionally, the inferred posterior distribution is used to simulate light curves, further evaluating the parameter retrieval performance against real data.

\subsection{AT2017gfo Parameter Retrieval}

\begin{figure}
     \centering
     \includegraphics[width=\linewidth]{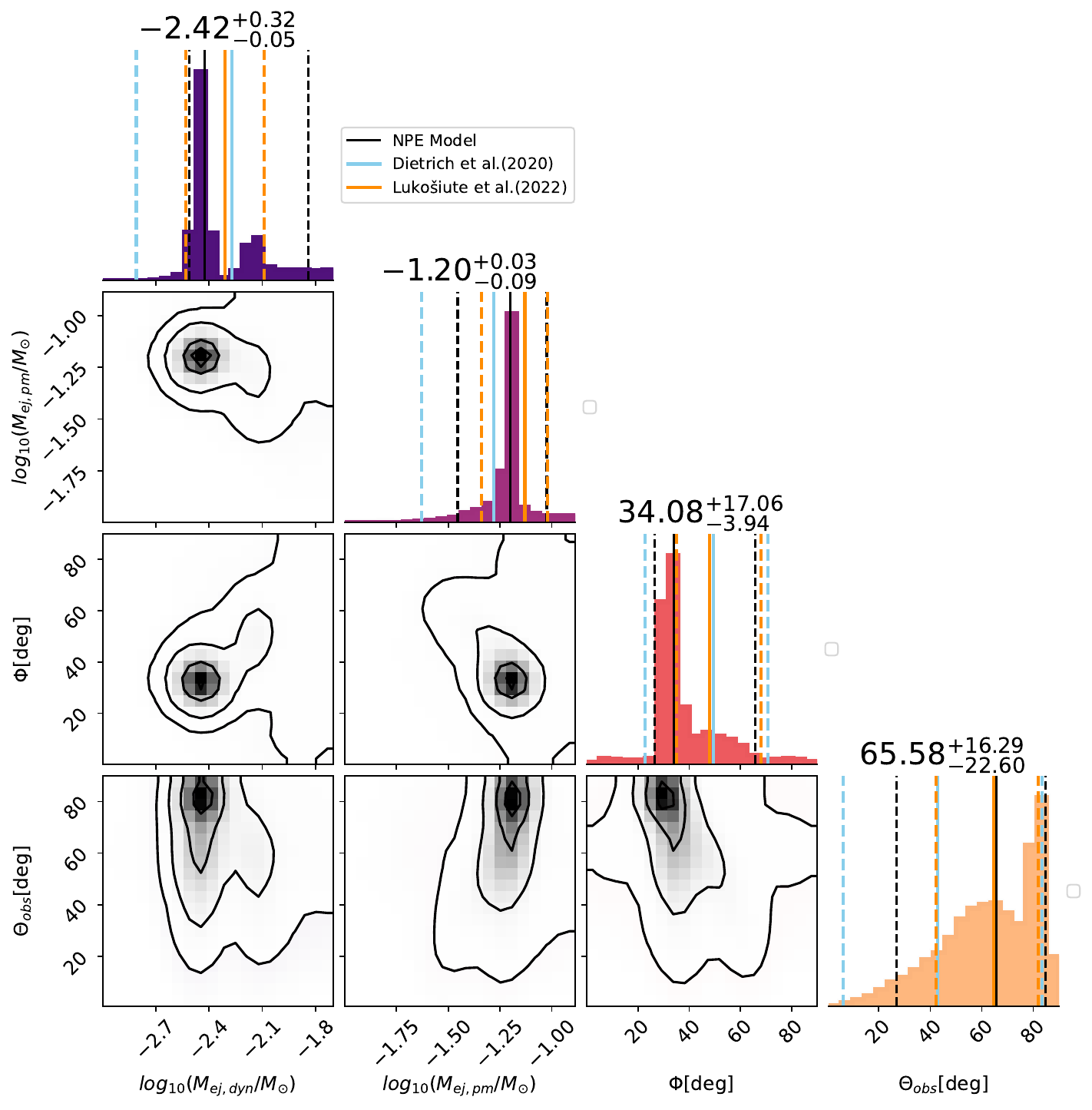}
     \caption{Corner plot of the inferred kilonova parameters for the AT\,2017gfo spectra at 10\%, 32\%, 68\% and 95\% confidence intervals. The median and 90\% interval are shown in vertical solid and dashed lines, respectively, and reported above each column. Results from \citet{Dietrich_2020sim} (light blue) and \citet{Lukosiute_2022} (orange) are also shown for comparison. We find excellent agreement between the median values of parameters inferred by different methods.}
     \label{fig:corner_real}
 \end{figure}

Our amortized inference procedure takes less than three seconds to infer the ensemble posterior distribution, which is the stack of the individual posteriors.  The corner plot (Figure~\ref{fig:corner_real}) shows one- and two-dimensional marginal ensemble posterior distributions for the  $M_{\rm ej,dyn}$, $M_{\rm ej,pm}$, $\phi$, $\cos(\theta_{\rm obs})$. We also present the corresponding optimal values obtained by \cite{Dietrich_2020sim} (light blue) and \cite{Lukosiute_2022} (orange).  The most probable values extracted via NPE are represented by black solid lines, while the $90\%$ interval is indicated by black dashed lines. Some intervals are beyond the plot range, and are omitted for the sake of clarity. 
The inferred posterior distributions are consistent with previous work, as our median values fall within the reported $90\%$ confidence interval by \cite{Dietrich_2020sim} and \cite{Lukosiute_2022}.
Furthermore, the median values from these sources are encompassed within the 1-$\sigma$ of our credible region, demonstrating a clear overlap between our analysis, even across different Bayesian inference methodologies. \cite{Dietrich_2020sim}, same kilonova model, uses light curve data, and incorporates priors coming from Gravitational waves and pulsar observations; \cite{Lukosiute_2022} also used \texttt{KilonovaNet}, but with a dynamic nested-sampling algorithm in a likelihood-based approach with light curve data.  The median values and the 90\% credible intervals are reported in Table \ref{tab:parameters}, We observe accurate recovery of kilonova parameters, particularly for the ejecta mass, demonstrating remarkable precision compared to likelihood-based approaches. 

\begin{deluxetable*}{ccccc}
\tablewidth{0pc}
\tabletypesize{\small}
\tablecaption{Kilonova Parameters\label{tab:parameters}}
\tablehead{\colhead{Model} &
\colhead{$\log(M_{\rm ej, dyn}/M_{\odot})$} &
\colhead{$\log(M_{\rm ej, pm}/M_{\odot})$} &
\colhead{$\phi$} &
\colhead{$\theta_{\rm obs}$} \\
&
&
&
[deg] &
[deg]
}
\startdata
\cite{Dietrich_2020sim} & $-2.27_{-0.54}^{+1.01}$ &  $-1.28_{-0.42}^{+0.35}$ & $49.50_{-26.65}^{+21.16}$ & $42.80_{-36.51}^{+40.62}$ \\ \hline 
\cite{Lukosiute_2022} &  $-2.31_{-0.22}^{+0.22}$ &  $-1.13_{-0.21}^{+0.11}$ &  $47.98_{-12.90}^{+20.21}$ & $64.55_{-22.34}^{+17.39}$ \\ \hline 
ANPE &  $-2.42_{-0.50}^{+0.32}$ &  $-1.20_{-0.09}^{+0.03}$ &  $34.08_{-3.94}^{+17.06}$ & $65.58_{-22.60}^{+16.29}$
\enddata
\tablecomments{Median values and $90\%$ credible intervals for kilonova parameters reported in various studies.}
\end{deluxetable*}

Our constraints on the parameter $\cos(\theta_{\rm obs})$ are the weakest, as discussed in Section~\ref{sec:kntestresults}. Importantly, this limitation is not exclusive to the NPE method;  \cite{Lukosiute_2022} and \cite{Dietrich_2020sim} similarly faces challenges in constraining the viewing angle in a likelihood-based approach. This outcome suggests that our NPE approach is capable of producing reliable posterior distribution on the real event. We can validate this assertion by visually comparing the real spectra with the reconstructed spectra.

\begin{figure*}
     \centering
     \includegraphics[width=1\linewidth]{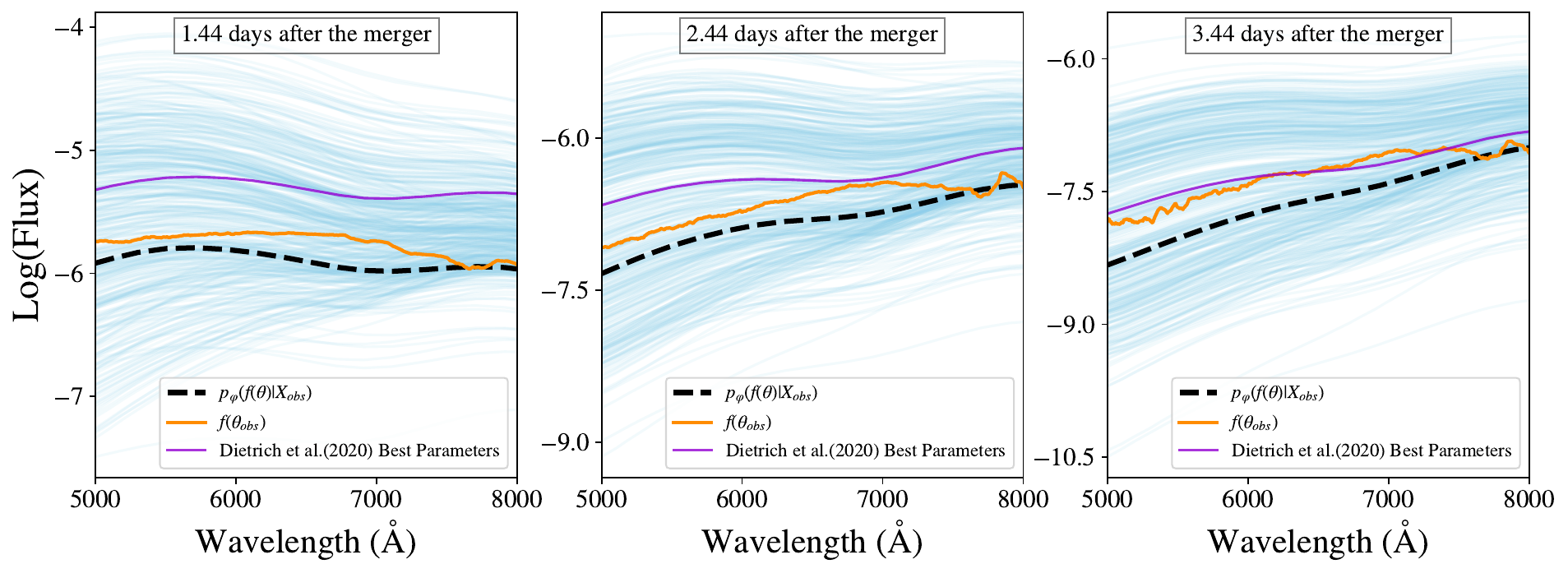}
     \caption{Spectroscopic time series of AT2017gfo (orange solid line), the spectra generated by {\tt KilonovaNet} using the Best fit of our ANPE model ( $p_\phi(f(\theta) \vert X_{\rm obs})$, black dashed line), plus spectra generated by 500 parameters set randomly sampled from the inferred posterior distribution (light blue solid line), and the spectra generated the best parameters from \citet{Dietrich_2020sim}, the vertical axis is Logarithm of the observed flux. }
     \label{fig:bestfit}
 \end{figure*}

Figure~\ref{fig:bestfit} displays a comparison of spectral data reconstructed using \texttt{KilonovaNet} surrogate model with our inferred posterior distributions ($NPE(f(\theta)|x_{\rm obs})$), \cite{Dietrich_2020sim}'s best parameters and the real spectra used as input to our model ($f(\theta_{\rm obs})$). We also display 500 sampled SEDs from the posterior distribution (light blue solid lines). The inferred spectra fit closely to the observable spectra, and to the blue reconstructed spectra. It shows that our model not only retrieve the parameters but also produces spectra close to the observable.

\begin{figure}
     \centering
     \includegraphics[width=1\linewidth]{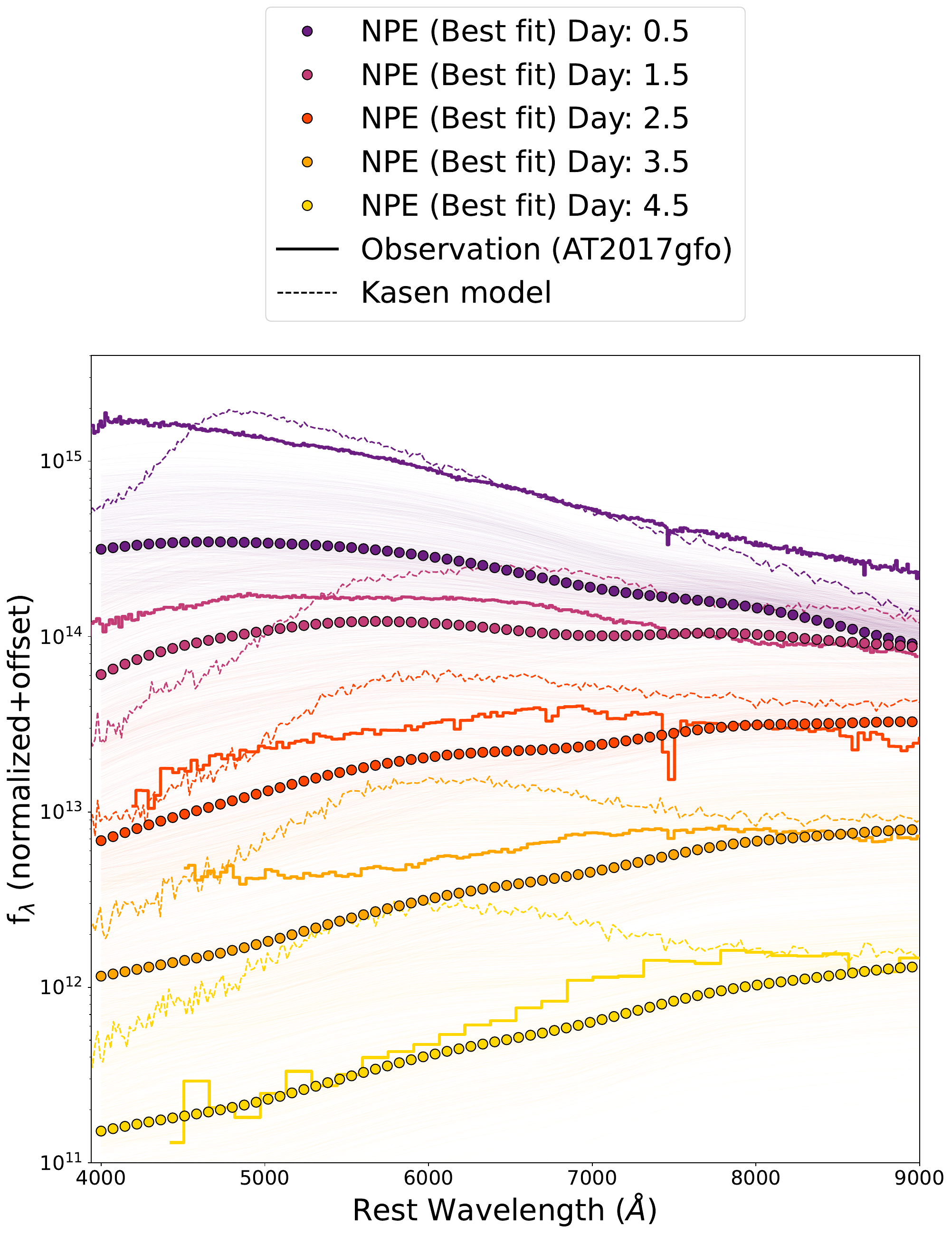}
     \caption{Spectroscopic time series of AT2017gfo (solid line) for five days after the merger, the spectra generated by {\tt KilonovaNet} using the Best fit of our ANPE model (dotted line), plus spectra generated by 300 parameters set randomly sampled from the inferred posterior distribution (solid thin line), and the Kasen model best fit made by \citet{Kilpa_Shappee_2017} (dashed line), the vertical axis is Logarithm of the observed flux.  }
     \label{fig:spec}
 \end{figure}

Additionally, to extend our analysis on the real data,  Figure~\ref{fig:spec} displays the fitted spectra from \cite{Kilpa_Shappee_2017} using \cite{Kasen_2017} model and the spectra reconstructed using our inferred posterior distribution, for five different times after the merger. Our model reconstructs the overall shape of the spectra from AT\,2017gfo for all times beyond 0.5 days post-merger with reasonable accuracy. 
The least favorable outcome is observed for the first 0.5 days post-merger, a time point for which the model was not specifically trained with. Nevertheless, it is crucial to acknowledge that the reported error for the early days ($<$0.5~days) generated by  \texttt{KilonovaNet} is substantial, potentially contributing to the disparity between the predicted and observed spectra.  These results visually confirms that our parameters estimation is not random or inaccurate at some extension. It produces spectra close to the real data used from training and even can be extended to days that were not given to the model.

\begin{figure*}
     \centering
     \includegraphics[width=\linewidth]{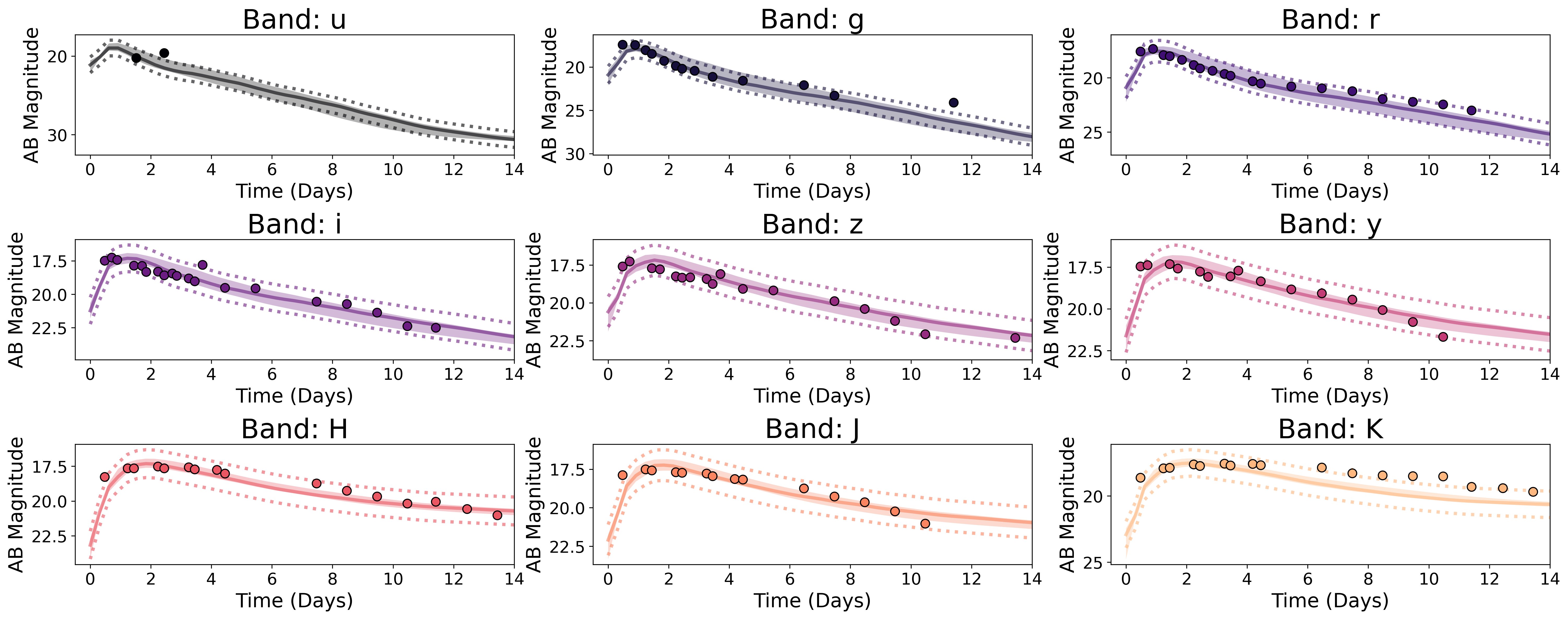}
     \caption{Light curves for AT\,2017gfo, showcasing observed values (points) and the predictions based on inferred parameters using the NPE model (solid lines). The shaded bands represent the 90\% confidence interval of light curves constructed from the posterior samples. Dashed lines indicate the 1 mag tolerance as reported by \citet{Lukosiute_2022}.}
     \label{fig:light_curve}
\end{figure*}

Lastly, we can also use \texttt{KilonovaNet} to generate light curves at different bands up to 14 days after the merger and compare it to the real data.
Figure~\ref{fig:light_curve} shows the light curves produced using our NPE method for different bands. Solid lines represent the inferred light curve using the best-fit parameters, and the shaded region the reconstructed light curves using the $90\%$ confidence interval from the inferred parameter distributions. Our model is able to accurately predict the overall shape of the light curves for all filters up to 14 days after the merger, considering a tolerance of one magnitude as reported in \cite{Lukosiute_2022}. 

\vspace{2cm}

\subsection{Limitations of these Models}\label{sec:caveats}

In the previous section, we demonstrated the accuracy of our NPE model for reproducing observables from kilonova spectra. Nevertheless, an important factor for the fidelity of our method, or any deep learning model, is the quality of the training data set, and, thus, the simulator and pre-processing steps used to generate the data. Below, we discuss the caveats and limitations of our model.

First, for our noise model, we assign uncertainties to noiseless photometric fluxes, by adding a Gaussian noise corresponding to 10\% of the flux at each point. We did not observe a notable enhancement with the inclusion of alternative noise levels.  This simplistic approach could be improved by adding realistic noise from instruments. Despite the lack of complex noise modelling, our ANPE model demonstrates remarkable agreement with likelihood-based methodologies.

In the context of the {\tt KilonovaNet} simulator, it is imperative to acknowledge its inherent limitations.  Generating synthetic observables is challenging, most radiative transfer simulations for kilonovae use simplified ejecta models, such as with analytical density structures \citep{Kasen_2017,bulla2019,Dietrich_2020sim}. Moreover, these simulations employ simplified opacity treatments, typically by combining line transitions into wavelength bins rather than calculating individual line-by-line opacities.

While our model demonstrates proficiency in accurately inferring kilonova parameters and replicating observables from the GW170817 event. Spectrographs equipped to capture high-resolution data may introduce additional spectral features, thus it may result in a more significant model misspecification. This results in a greater disparity between simulated and actual data. Therefore, we highlight the limitation of the simulator on generating realistic spectral features and noise. This supports the necessity for a more sophisticated radiative transfer simulator capable of generating spectral data that incorporates fluorescence effects and associate spectral features with the emitting and absorbing lines of individual elements \citep{Shingles_2023}, potentially enhancing the constraints on the observing viewing angle.

\section{Conclusion}\label{sec:conclusion}

This study presents a novel application of machine learning to astrophysics, which combines radiative transfer simulation model with simulation-based inference (SBI) applied to spectral data for the first time. 
By demonstrating its effectiveness on both synthetic data and the GW170817 event, we showcased the model's ability to accurately retrieve kilonova parameters while offering significant computational advantages over traditional methods.

One of the main advantages of neural posterior estimation is its amortization of the inference procedure. Once trained, the inference process no longer depends on on-the-fly simulations and can be repeated multiple times with different observations at very low computational cost. Given the speed with which the ANPE model performs during parameter inference, it can retrieve kilonova parameters for large datasets. Therefore, It can be tested by numerous diagnostics such as Coverage diagnostic, SBC and PPC, which would be impractical or extremely time-consuming for MCMC methods. The speed-up provided in the parameter inference and spectra retrieval also enables the exploration of multiple simulation models, such as the one proposed by  \citet{Kasen_2017}, over limited observations in a reasonable time. Additionally, the derived posterior from ANPE can serve as valuable prior information for likelihood-based approaches, further accelerating the overall inference time.

Unlike traditional methods like MCMC and nested sampling,  NPE is capable of targeting the 1-d or 2-d marginal posterior distributions, rather than the full joint posterior. This is particularly crucial for upcoming radiative transfer simulation models that are anticipated to incorporate numerous nuisance parameters.

Another critical driver of these algorithms is the rapidly increasing capabilities of machine learning, which enable us to analyze high-dimensional data efficiently and automatically extract features from the data, allowing for faster inference.  Through our demonstration, we showcase our model's ability to accurately replicate both the spectra and light curves of synthetic data and the AT,2017gfo event. The demonstrated accuracy and efficiency of ANPE highlight the potential of SBI in revolutionizing parameter estimation within astrophysical simulations.

 In summary, our exploration into the merger of machine learning and physics, specifically applied to kilonova events, has yielded promising results. We demonstrate in this work that ANPE provides an alternative approach for Bayesian inference in SED modeling when applied to synthetic and real data, and this represents a significant stride forward in the study of these explosive phenomena.

\section*{Acknowledgments}

CRB acknowledges the financial support from CNPq (316072/2021-4) and from FAPERJ (grants 201.456/2022 and 210.330/2022) and the FINEP contract 01.22.0505.00 (ref. 1891/22). C.D.K. acknowledges partial support from a CIERA postdoctoral fellowship. The authors made use of Sci-Mind servers machines developed by the CBPF AI LAB team and would like to thank A. Santos, P. Russano, and M. Portes de Albuquerque for all the support in infrastructure matters.


\section*{Data Availability}  

The code and dataset used to perform the experiments presented in this paper is openly available in our GitHub repository: \href{https://github.com/phelipedarc/Kilonova_Spectral_Modelling_NPE.git}{Kilonova Spectra Modelling}

\bibliography{step}

\begin{thebibliography}{}
\expandafter\ifx\csname natexlab\endcsname\relax\def\natexlab#1{#1}\fi
\providecommand{\url}[1]{\href{#1}{#1}}
\providecommand{\dodoi}[1]{doi:~\href{http://doi.org/#1}{\nolinkurl{#1}}}
\providecommand{\doeprint}[1]{\href{http://ascl.net/#1}{\nolinkurl{http://ascl.net/#1}}}
\providecommand{\doarXiv}[1]{\href{https://arxiv.org/abs/#1}{\nolinkurl{https://arxiv.org/abs/#1}}}

\bibitem[{Abbott {et~al.}(2017)Abbott, Abbott, Abbott, Abernathy, Acernese, Ackley, Adams, Adams, Addesso, Adhikari, {et~al.}}]{abbott2017search}
Abbott, B.~P., Abbott, R., Abbott, T.~D., {et~al.} 2017, The Astrophysical Journal, 841, 89

\bibitem[{{Abbott} {et~al.}(2017){Abbott}, {Abbott}, {Abbott}, {Acernese}, {Ackley}, {Adams}, {Adams}, {Addesso}, {Adhikari}, {Adya}, \& et~al.}]{2017Natur.551...85A}
{Abbott}, B.~P., {Abbott}, R., {Abbott}, T.~D., {et~al.} 2017, Nature, 551, 85, \dodoi{10.1038/nature24471}

\bibitem[{Abbott {et~al.}(2020)Abbott, Abbott, Abbott, Abraham, Acernese, Ackley, Adams, Adya, Affeldt, Agathos, {et~al.}}]{abbott2020prospects}
Abbott, B.~P., Abbott, R., Abbott, T., {et~al.} 2020, Living reviews in relativity, 23, 1

\bibitem[{Acernese {et~al.}(2014)Acernese, Agathos, Agatsuma, Aisa, Allemandou, Allocca, Amarni, Astone, Balestri, Ballardin, Barone, Baronick, Barsuglia, Basti, Basti, Bauer, Bavigadda, Bejger, Beker, Belczynski, Bersanetti, Bertolini, Bitossi, Bizouard, Bloemen, Blom, Boer, Bogaert, Bondi, Bondu, Bonelli, Bonnand, Boschi, Bosi, Bouedo, Bradaschia, Branchesi, Briant, Brillet, Brisson, Bulik, Bulten, Buskulic, Buy, Cagnoli, Calloni, Campeggi, Canuel, Carbognani, Cavalier, Cavalieri, Cella, Cesarini, Chassande-Mottin, Chincarini, Chiummo, Chua, Cleva, Coccia, Cohadon, Colla, Colombini, Conte, Coulon, Cuoco, Dalmaz, D’Antonio, Dattilo, Davier, Day, Debreczeni, Degallaix, Deléglise, Pozzo, Dereli, Rosa, Fiore, Lieto, Virgilio, Doets, Dolique, Drago, Ducrot, Endrőczi, Fafone, Farinon, Ferrante, Ferrini, Fidecaro, Fiori, Flaminio, Fournier, Franco, Frasca, Frasconi, Gammaitoni, Garufi, Gaspard, Gatto, Gemme, Gendre, Genin, Gennai, Ghosh, Giacobone, Giazotto, Gouaty, Granata, Greco, Groot, Guidi, Harms,
  Heidmann, Heitmann, Hello, Hemming, Hennes, Hofman, Jaranowski, Jonker, Kasprzack, Kéfélian, Kowalska, Kraan, Królak, Kutynia, Lazzaro, Leonardi, Leroy, Letendre, Li, Lieunard, Lorenzini, Loriette, Losurdo, Magazzù, Majorana, Maksimovic, Malvezzi, Man, Mangano, Mantovani, Marchesoni, Marion, Marque, Martelli, Martellini, Masserot, Meacher, Meidam, Mezzani, Michel, Milano, Minenkov, Moggi, Mohan, Montani, Morgado, Mours, Mul, Nagy, Nardecchia, Naticchioni, Nelemans, Neri, Neri, Nocera, Pacaud, Palomba, Paoletti, Paoli, Pasqualetti, Passaquieti, Passuello, Perciballi, Petit, Pichot, Piergiovanni, Pillant, Piluso, Pinard, Poggiani, Prijatelj, Prodi, Punturo, Puppo, Rabeling, Rácz, Rapagnani, Razzano, Re, Regimbau, Ricci, Robinet, Rocchi, Rolland, Romano, Rosińska, Ruggi, Saracco, Sassolas, Schimmel, Sentenac, Sequino, Shah, Siellez, Straniero, Swinkels, Tacca, Tonelli, Travasso, Turconi, Vajente, van Bakel, van Beuzekom, van~den Brand, Broeck, van~der Sluys, van Heijningen, Vasúth, Vedovato, Veitch,
  Verkindt, Vetrano, Viceré, Vinet, Visser, Vocca, Ward, Was, Wei, Yvert, żny, \& Zendri}]{Acernese_2015}
Acernese, F., Agathos, M., Agatsuma, K., {et~al.} 2014, Classical and Quantum Gravity, 32, 024001, \dodoi{10.1088/0264-9381/32/2/024001}

\bibitem[{{Alexander} {et~al.}(2017){Alexander}, {Berger}, {Fong}, {Williams}, {Guidorzi}, {Margutti}, {Metzger}, {Annis}, {Blanchard}, {Brout}, {Brown}, {Chen}, {Chornock}, {Cowperthwaite}, {Drout}, {Eftekhari}, {Frieman}, {Holz}, {Nicholl}, {Rest}, {Sako}, {Soares-Santos}, \& {Villar}}]{2017alex}
{Alexander}, K.~D., {Berger}, E., {Fong}, W., {et~al.} 2017, \apjl, 848, L21, \dodoi{10.3847/2041-8213/aa905d}

\bibitem[{{Alfradique} {et~al.}(2024){Alfradique}, {Bom}, {Palmese}, {Teixeira}, {Santana-Silva}, {Drlica-Wagner}, {Riley}, {Mart{\'\i}nez-V{\'a}zquez}, {Sand}, {Stringfellow}, {Medina}, {Carballo-Bello}, {Choi}, {Esteves}, {Limberg}, {Mutlu-Pakdil}, {No{\"e}l}, {Pace}, {Sakowska}, \& {Wu}}]{alfradique24}
{Alfradique}, V., {Bom}, C.~R., {Palmese}, A., {et~al.} 2024, \mnras, 528, 3249, \dodoi{10.1093/mnras/stae086}

\bibitem[{Anand {et~al.}(2020)Anand, Coughlin, Kasliwal, Bulla, Ahumada, Carracedo, Almualla, Andreoni, Stein, Foucart, Singer, Sollerman, Bellm, Bolin, Caballero-Garc{\'{\i}}a, Castro-Tirado, Cenko, De, Dekany, Duev, Feeney, Fremling, Goldstein, Golkhou, Graham, Guessoum, Hankins, Hu, Kong, Kool, Kulkarni, Kumar, Laher, Masci, Mr{\'{o}}z, Nissanke, Porter, Reusch, Riddle, Rosnet, Rusholme, Serabyn, S{\'{a}}nchez-Ram{\'{\i}}rez, Rigault, Shupe, Smith, Soumagnac, Walters, \& Valeev}]{Anand_2020}
Anand, S., Coughlin, M.~W., Kasliwal, M.~M., {et~al.} 2020, Nature Astronomy, 5, 46, \dodoi{10.1038/s41550-020-1183-3}

\bibitem[{{Andreoni} {et~al.}(2024){Andreoni}, {Coughlin}, {Criswell}, {Bulla}, {Toivonen}, {Singer}, {Palmese}, {Burns}, {Gezari}, {Kasliwal}, {Kiendrebeogo}, {Mahabal}, {Moriya}, {Rest}, {Scolnic}, {Simcoe}, {Soon}, {Stein}, \& {Travouillon}}]{Andreoni24}
{Andreoni}, I., {Coughlin}, M.~W., {Criswell}, A.~W., {et~al.} 2024, Astroparticle Physics, 155, 102904, \dodoi{10.1016/j.astropartphys.2023.102904}

\bibitem[{Bom {et~al.}(2024)Bom, Annis, Garcia, Palmese, Sherman, Soares-Santos, Santana-Silva, Morgan, Bechtol, Davis, Diehl, Allam, Bachmann, Fraga, García-Bellido, Gill, Herner, Kilpatrick, Makler, E., Pereira, Pineda, Santos, Tucker, Wiesner, Aguena, Alves, Bacon, Bernardinelli, Bertin, Bocquet, Brooks, Kind, Carretero, Conselice, Costanzi, da~Costa, Vicente, Desai, Doel, Everett, Ferrero, Frieman, Gatti, Gerdes, Gruen, Gruendl, Gutierrez, Hinton, Hollowood, Honscheid, James, Kuehn, Kuropatkin, Melchior, Mena-Fernández, Menanteau, Pieres, Malagón, Raveri, Rodriguez-Monroy, Sanchez, Santiago, Sevilla-Noarbe, Smith, Suchyta, Swanson, Tarle, To, \& Weaverdyck}]{Bom_2024}
Bom, C.~R., Annis, J., Garcia, A., {et~al.} 2024, The Astrophysical Journal, 960, 122, \dodoi{10.3847/1538-4357/ad0462}

\bibitem[{Bulla(2019)}]{bulla2019}
Bulla, M. 2019, Monthly Notices of the Royal Astronomical Society, 489, 5037, \dodoi{10.1093/mnras/stz2495}

\bibitem[{Cole {et~al.}(2022)Cole, Miller, Witte, Cai, Grootes, Nattino, \& Weniger}]{Cole_2022}
Cole, A., Miller, B.~K., Witte, S.~J., {et~al.} 2022, Journal of Cosmology and Astroparticle Physics, 2022, 004, \dodoi{10.1088/1475-7516/2022/09/004}

\bibitem[{Collaboration {et~al.}(2015)Collaboration, Aasi, Abbott, Abbott, Abbott, Abernathy, Ackley, Adams, Adams, Addesso, Adhikari, Adya, Affeldt, Aggarwal, Aguiar, Ain, Ajith, Alemic, Allen, Amariutei, Anderson, Anderson, Arai, Araya, Arceneaux, Areeda, Ashton, Ast, Aston, Aufmuth, Aulbert, Aylott, Babak, Baker, Ballmer, Barayoga, Barbet, Barclay, Barish, Barker, Barr, Barsotti, Bartlett, Barton, Bartos, Bassiri, Batch, Baune, Behnke, Bell, Bell, Benacquista, Bergman, Bergmann, Berry, Betzwieser, Bhagwat, Bhandare, Bilenko, Billingsley, Birch, Biscans, Biwer, Blackburn, Blackburn, Blair, Blair, Bock, Bodiya, Bojtos, Bond, Bork, Born, Bose, Brady, Braginsky, Brau, Bridges, Brinkmann, Brooks, Brown, Brown, Brown, Buchman, Buikema, Buonanno, Cadonati, Bustillo, Camp, Cannon, Cao, Capano, Caride, Caudill, Cavaglià, Cepeda, Chakraborty, Chalermsongsak, Chamberlin, Chao, Charlton, Chen, Cho, Cho, Chow, Christensen, Chu, Chung, Ciani, Clara, Clark, Collette, Cominsky, Constancio, Cook, Corbitt, Cornish, Corsi,
  Costa, Coughlin, Countryman, Couvares, Coward, Cowart, Coyne, Coyne, Craig, Creighton, Creighton, Cripe, Crowder, Cumming, Cunningham, Cutler, Dahl, Canton, Damjanic, Danilishin, Danzmann, Dartez, Dave, Daveloza, Davies, Daw, DeBra, Pozzo, Denker, Dent, Dergachev, DeRosa, DeSalvo, Dhurandhar, D´ıaz, Palma, Dojcinoski, Dominguez, Donovan, Dooley, Doravari, Douglas, Downes, Driggers, Du, Dwyer, Eberle, Edo, Edwards, Edwards, Effler, Eggenstein, Ehrens, Eichholz, Eikenberry, Essick, Etzel, Evans, Evans, Factourovich, Fairhurst, Fan, Fang, Farr, Farr, Favata, Fays, Fehrmann, Fejer, Feldbaum, Ferreira, Fisher, Frei, Freise, Frey, Fricke, Fritschel, Frolov, Fuentes-Tapia, Fulda, Fyffe, Gair, Gaonkar, Gehrels, Gergely´, Giaime, Giardina, Gleason, Goetz, Goetz, Gondan, González, Gordon, Gorodetsky, Gossan, Goßler, Gräf, Graff, Grant, Gras, Gray, Greenhalgh, Gretarsson, Grote, Grunewald, Guido, Guo, Gushwa, Gustafson, Gustafson, Hacker, Hall, Hammond, Hanke, Hanks, Hanna, Hannam, Hanson, Hardwick, Harry,
  Harry, Hart, Hartman, Haster, Haughian, Hee, Heintze, Heinzel, Hendry, Heng, Heptonstall, Heurs, Hewitson, Hild, Hoak, Hodge, Hollitt, Holt, Hopkins, Hosken, Hough, Houston, Howell, Hu, Huerta, Hughey, Husa, Huttner, Huynh, Huynh-Dinh, Idrisy, Indik, Ingram, Inta, Islas, Isler, Isogai, Iyer, Izumi, Jacobson, Jang, Jawahar, Ji, Jiménez-Forteza, Johnson, Jones, Jones, Ju, Haris, Kalogera, Kandhasamy, Kang, Kanner, Katsavounidis, Katzman, Kaufer, Kaufer, Kaur, Kawabe, Kawazoe, Keiser, Keitel, Kelley, Kells, Keppel, Key, Khalaidovski, Khalili, Khazanov, Kim, Kim, Kim, Kim, Kim, King, King, Kinzel, Kissel, Klimenko, Kline, Koehlenbeck, Kokeyama, Kondrashov, Korobko, Korth, Kozak, Kringel, Krishnan, Krueger, Kuehn, Kumar, Kumar, Kuo, Landry, Lantz, Larson, Lasky, Lazzarini, Lazzaro, Le, Leaci, Leavey, Lebigot, Lee, Lee, Lee, Leong, Levin, Levine, Lewis, Li, Libbrecht, Libson, Lin, Littenberg, Lockerbie, Lockett, Logue, Lombardi, Lormand, Lough, Lubinski, Lück, Lundgren, Lynch, Ma, Macarthur, MacDonald,
  Machenschalk, MacInnis, Macleod, Magaña-Sandoval, Magee, Mageswaran, Maglione, Mailand, Mandel, Mandic, Mangano, Mansell, Márka, Márka, Markosyan, Maros, Martin, Martin, Martynov, Marx, Mason, Massinger, Matichard, Matone, Mavalvala, Mazumder, Mazzolo, McCarthy, McClelland, McCormick, McGuire, McIntyre, McIver, McLin, McWilliams, Meadors, Meinders, Melatos, Mendell, Mercer, Meshkov, Messenger, Meyers, Miao, Middleton, Mikhailov, Miller, Miller, Millhouse, Ming, Mirshekari, Mishra, Mitra, Mitrofanov, Mitselmakher, Mittleman, Moe, Mohanty, Mohapatra, Moore, Moraru, Moreno, Morriss, Mossavi, Mow-Lowry, Mueller, Mueller, Mukherjee, Mullavey, Munch, Murphy, Murray, Mytidis, Nash, Nayak, Necula, Nedkova, Newton, Nguyen, Nielsen, Nissanke, Nitz, Nolting, Normandin, Nuttall, Ochsner, O’Dell, Oelker, Ogin, Oh, Oh, Ohme, Oppermann, Oram, O’Reilly, Ortega, O’Shaughnessy, Osthelder, Ott, Ottaway, Ottens, Overmier, Owen, Padilla, Pai, Pai, Palashov, Pal-Singh, Pan, Pankow, Pannarale, Pant, Papa, Paris, Patrick,
  Pedraza, Pekowsky, Pele, Penn, Perreca, Phelps, Pierro, Pinto, Pitkin, Poeld, Post, Poteomkin, Powell, Prasad, Predoi, Premachandra, Prestegard, Price, Principe, Privitera, Prix, Prokhorov, Puncken, Pürrer, Qin, Quetschke, Quintero, Quiroga, Quitzow-James, Raab, Rabeling, Radkins, Raffai, Raja, Rajalakshmi, Rakhmanov, Ramirez, Raymond, Reed, Reid, Reitze, Reula, Riles, Robertson, Robie, Rollins, Roma, Romano, Romanov, Romie, Rowan, Rüdiger, Ryan, Sachdev, Sadecki, Sadeghian, Saleem, Salemi, Sammut, Sandberg, Sanders, Sannibale, Santiago-Prieto, Sathyaprakash, Saulson, Savage, Sawadsky, Scheuer, Schilling, Schmidt, Schnabel, Schofield, Schreiber, Schuette, Schutz, Scott, Scott, Sellers, Sengupta, Sergeev, Serna, Sevigny, Shaddock, Shahriar, Shaltev, Shao, Shapiro, Shawhan, Shoemaker, Sidery, Siemens, Sigg, Silva, Simakov, Singer, Singer, Singh, Sintes, Slagmolen, Smith, Smith, Smith, Smith-Lefebvre, Son, Sorazu, Souradeep, Staley, Stebbins, Steinke, Steinlechner, Steinlechner, Steinmeyer, Stephens,
  Steplewski, Stevenson, Stone, Strain, Strigin, Sturani, Stuver, Summerscales, Sutton, Szczepanczyk, Szeifert, Talukder, Tanner, Tápai, Tarabrin, Taracchini, Taylor, Tellez, Theeg, Thirugnanasambandam, Thomas, Thomas, Thorne, Thorne, Thrane, Tiwari, Tomlinson, Torres, Torrie, Traylor, Tse, Tshilumba, Ugolini, Unnikrishnan, Urban, Usman, Vahlbruch, Vajente, Valdes, Vallisneri, van Veggel, Vass, Vaulin, Vecchio, Veitch, Veitch, Venkateswara, Vincent-Finley, Vitale, Vo, Vorvick, Vousden, Vyatchanin, Wade, Wade, Wade, Walker, Wallace, Walsh, Wang, Wang, Wang, Ward, Warner, Was, Weaver, Weinert, Weinstein, Weiss, Welborn, Wen, Wessels, Westphal, Wette, Whelan, Whitcomb, White, Whiting, Wilkinson, Williams, Williams, Williamson, Willis, Willke, Wimmer, Winkler, Wipf, Wittel, Woan, Worden, Xie, Yablon, Yakushin, Yam, Yamamoto, Yancey, Yang, Zanolin, Zhang, Zhang, Zhang, Zhang, Zhao, Zhou, Zhu, Zucker, Zuraw, \& Zweizig}]{Aasi_2015}
Collaboration, T. L.~S., Aasi, J., Abbott, B.~P., {et~al.} 2015, Classical and Quantum Gravity, 32, 074001, \dodoi{10.1088/0264-9381/32/7/074001}

\bibitem[{{Coughlin} {et~al.}(2020){Coughlin}, {Antier}, {Dietrich}, {Foley}, {Heinzel}, {Bulla}, {Christensen}, {Coulter}, {Issa}, \& {Khetan}}]{hubble_kn}
{Coughlin}, M.~W., {Antier}, S., {Dietrich}, T., {et~al.} 2020, Nature Communications, 11, 4129, \dodoi{10.1038/s41467-020-17998-5}

\bibitem[{{Coulter} {et~al.}(2017){Coulter}, {Foley}, {Kilpatrick}, {Drout}, {Piro}, {Shappee}, {Siebert}, {Simon}, {Ulloa}, {Kasen}, {Madore}, {Murguia-Berthier}, {Pan}, {Prochaska}, {Ramirez-Ruiz}, {Rest}, \& {Rojas-Bravo}}]{Coulter17}
{Coulter}, D.~A., {Foley}, R.~J., {Kilpatrick}, C.~D., {et~al.} 2017, Science, 358, 1556, \dodoi{10.1126/science.aap9811}

\bibitem[{Cranmer {et~al.}(2020)Cranmer, Brehmer, \& Louppe}]{Cranmer_2020}
Cranmer, K., Brehmer, J., \& Louppe, G. 2020, Proceedings of the National Academy of Sciences, 117, 30055, \dodoi{10.1073/pnas.1912789117}

\bibitem[{Dalmasso {et~al.}(2020)Dalmasso, Pospisil, Lee, Izbicki, Freeman, \& Malz}]{Dalmasso_2020}
Dalmasso, N., Pospisil, T., Lee, A., {et~al.} 2020, Astronomy and Computing, 30, 100362, \dodoi{10.1016/j.ascom.2019.100362}

\bibitem[{Deistler {et~al.}(2022)Deistler, Goncalves, \& Macke}]{deistler2022truncated}
Deistler, M., Goncalves, P.~J., \& Macke, J.~H. 2022, Truncated proposals for scalable and hassle-free simulation-based inference.
\newblock \doarXiv{2210.04815}

\bibitem[{Dietrich {et~al.}(2020)Dietrich, Coughlin, Pang, Bulla, Heinzel, Issa, Tews, \& Antier}]{Dietrich_2020sim}
Dietrich, T., Coughlin, M.~W., Pang, P. T.~H., {et~al.} 2020, Science, 370, 1450, \dodoi{10.1126/science.abb4317}

\bibitem[{Dietrich \& Ujevic(2017)}]{Dietrich_2017modelying}
Dietrich, T., \& Ujevic, M. 2017, Classical and Quantum Gravity, 34, 105014, \dodoi{10.1088/1361-6382/aa6bb0}

\bibitem[{Gabry {et~al.}(2019)Gabry, Simpson, Vehtari, Betancourt, \& Gelman}]{Gabry_2019}
Gabry, J., Simpson, D., Vehtari, A., Betancourt, M., \& Gelman, A. 2019, Journal of the Royal Statistical Society Series A: Statistics in Society, 182, 389–402, \dodoi{10.1111/rssa.12378}

\bibitem[{Greenberg {et~al.}(2019)Greenberg, Nonnenmacher, \& Macke}]{snpe-apt}
Greenberg, D., Nonnenmacher, M., \& Macke, J. 2019, in Proceedings of Machine Learning Research, Vol.~97, Proceedings of the 36th International Conference on Machine Learning, ed. K.~Chaudhuri \& R.~Salakhutdinov (Long Beach, California, USA: PMLR), 2404--2414.
\newblock \url{http://proceedings.mlr.press/v97/greenberg19a.html}

\bibitem[{Hermans {et~al.}(2022)Hermans, Delaunoy, Rozet, Wehenkel, Begy, \& Louppe}]{hermans2022trust}
Hermans, J., Delaunoy, A., Rozet, F., {et~al.} 2022, A Trust Crisis In Simulation-Based Inference? Your Posterior Approximations Can Be Unfaithful.
\newblock \doarXiv{2110.06581}

\bibitem[{{Ivezi{\'c}} {et~al.}(2019){Ivezi{\'c}}, {Kahn}, {Tyson}, {Abel}, {Acosta}, {Allsman}, {Alonso}, {AlSayyad}, {Anderson}, {Andrew}, \& et~al.}]{verarubin2019ApJ...873..111I}
{Ivezi{\'c}}, {\v Z}., {Kahn}, S.~M., {Tyson}, J.~A., {et~al.} 2019, The Astrophysical Journal, 873, 111, \dodoi{10.3847/1538-4357/ab042c}

\bibitem[{Kasen {et~al.}(2017)Kasen, Metzger, Barnes, Quataert, \& Ramirez-Ruiz}]{Kasen_2017}
Kasen, D., Metzger, B., Barnes, J., Quataert, E., \& Ramirez-Ruiz, E. 2017, Nature, 551, 80, \dodoi{10.1038/nature24453}

\bibitem[{Kawaguchi {et~al.}(2020)Kawaguchi, Shibata, \& Tanaka}]{Kawaguchi_2020}
Kawaguchi, K., Shibata, M., \& Tanaka, M. 2020, The Astrophysical Journal, 889, 171, \dodoi{10.3847/1538-4357/ab61f6}

\bibitem[{Kingma \& Ba(2017)}]{kingma2017adam}
Kingma, D.~P., \& Ba, J. 2017, Adam: A Method for Stochastic Optimization.
\newblock \doarXiv{1412.6980}

\bibitem[{Kingma \& Welling(2022)}]{kingma2022autoencoding}
Kingma, D.~P., \& Welling, M. 2022, Auto-Encoding Variational Bayes.
\newblock \doarXiv{1312.6114}

\bibitem[{{Lattimer} \& {Schramm}(1974)}]{1974ApJ...192L.145L}
{Lattimer}, J.~M., \& {Schramm}, D.~N. 1974, \apjl, 192, L145, \dodoi{10.1086/181612}

\bibitem[{Levan {et~al.}(2024)}]{levan2023}
Levan, A.~J., {et~al.} 2024, Nature, 626, 737, \dodoi{10.1038/s41586-023-06759-1}

\bibitem[{Li \& Paczy{\'{n} }ski(1998)}]{Li_1998}
Li, L.-X., \& Paczy{\'{n} }ski, B. 1998, The Astrophysical Journal, 507, L59, \dodoi{10.1086/311680}

\bibitem[{Lopez-Paz \& Oquab(2018)}]{lopezpaz2018revisiting}
Lopez-Paz, D., \& Oquab, M. 2018, Revisiting Classifier Two-Sample Tests.
\newblock \doarXiv{1610.06545}

\bibitem[{Lueckmann {et~al.}(2021)Lueckmann, Boelts, Greenberg, Gonçalves, \& Macke}]{lueckmann2021benchmarking}
Lueckmann, J.-M., Boelts, J., Greenberg, D.~S., Gonçalves, P.~J., \& Macke, J.~H. 2021, Benchmarking Simulation-Based Inference.
\newblock \doarXiv{2101.04653}

\bibitem[{Luko{\v{s}}iute {et~al.}(2022)Luko{\v{s}}iute, Raaijmakers, Doctor, Soares-Santos, \& Nord}]{Lukosiute_2022}
Luko{\v{s}}iute, K., Raaijmakers, G., Doctor, Z., Soares-Santos, M., \& Nord, B. 2022, Monthly Notices of the Royal Astronomical Society, 516, 1137, \dodoi{10.1093/mnras/stac2342}

\bibitem[{Margalit \& Metzger(2017)}]{Margalit_2017}
Margalit, B., \& Metzger, B.~D. 2017, The Astrophysical Journal, 850, L19, \dodoi{10.3847/2041-8213/aa991c}

\bibitem[{Metzger(2019)}]{Metzger_2019}
Metzger, B.~D. 2019, Living Reviews in Relativity, 23, \dodoi{10.1007/s41114-019-0024-0}

\bibitem[{Montel {et~al.}(2023)Montel, Alvey, \& Weniger}]{montel2023scalable}
Montel, N.~A., Alvey, J., \& Weniger, C. 2023, Scalable inference with Autoregressive Neural Ratio Estimation.
\newblock \doarXiv{2308.08597}

\bibitem[{Papamakarios \& Murray(2018)}]{papamakarios2018fast}
Papamakarios, G., \& Murray, I. 2018, Fast $\epsilon$-free Inference of Simulation Models with Bayesian Conditional Density Estimation.
\newblock \doarXiv{1605.06376}

\bibitem[{Papamakarios {et~al.}(2021)Papamakarios, Nalisnick, Rezende, Mohamed, \& Lakshminarayanan}]{papamakarios2021normalizing}
Papamakarios, G., Nalisnick, E., Rezende, D.~J., Mohamed, S., \& Lakshminarayanan, B. 2021, The Journal of Machine Learning Research, 22, 2617

\bibitem[{Radice {et~al.}(2018)Radice, Perego, Zappa, \& Bernuzzi}]{Radice_2018}
Radice, D., Perego, A., Zappa, F., \& Bernuzzi, S. 2018, The Astrophysical Journal Letters, 852, L29, \dodoi{10.3847/2041-8213/aaa402}

\bibitem[{{Rastinejad} {et~al.}(2021){Rastinejad}, {Fong}, {Kilpatrick}, {Paterson}, {Tanvir}, {Levan}, {Metzger}, {Berger}, {Chornock}, {Cobb}, {Laskar}, {Milne}, {Nugent}, \& {Smith}}]{Rastinejad21}
{Rastinejad}, J.~C., {Fong}, W., {Kilpatrick}, C.~D., {et~al.} 2021, \apj, 916, 89, \dodoi{10.3847/1538-4357/ac04b4}

\bibitem[{{Rastinejad} {et~al.}(2022){Rastinejad}, {Gompertz}, {Levan}, {Fong}, {Nicholl}, {Lamb}, {Malesani}, {Nugent}, {Oates}, {Tanvir}, {de Ugarte Postigo}, {Kilpatrick}, {Moore}, {Metzger}, {Ravasio}, {Rossi}, {Schroeder}, {Jencson}, {Sand}, {Smith}, {Ag{\"u}{\'\i} Fern{\'a}ndez}, {Berger}, {Blanchard}, {Chornock}, {Cobb}, {De Pasquale}, {Fynbo}, {Izzo}, {Kann}, {Laskar}, {Marini}, {Paterson}, {Escorial}, {Sears}, \& {Th{\"o}ne}}]{Rastinejad22}
{Rastinejad}, J.~C., {Gompertz}, B.~P., {Levan}, A.~J., {et~al.} 2022, nature, 612, 223, \dodoi{10.1038/s41586-022-05390-w}

\bibitem[{Shappee {et~al.}(2017)Shappee, Simon, Drout, Piro, Morrell, Prieto, Kasen, Holoien, Kollmeier, Kelson, Coulter, Foley, Kilpatrick, Siebert, Madore, Murguia-Berthier, Pan, Prochaska, Ramirez-Ruiz, Rest, Adams, Alatalo, Ba{\~{n} }ados, Baughman, Bernstein, Bitsakis, Boutsia, Bravo, Mille, Higgs, Ji, Maravelias, Marshall, Placco, Prieto, \& Wan}]{Kilpa_Shappee_2017}
Shappee, B.~J., Simon, J.~D., Drout, M.~R., {et~al.} 2017, Science, 358, 1574, \dodoi{10.1126/science.aaq0186}

\bibitem[{Shingles {et~al.}(2023)Shingles, Collins, Vijayan, Flörs, Just, Leck, Xiong, Bauswein, Martínez-Pinedo, \& Sim}]{Shingles_2023}
Shingles, L.~J., Collins, C.~E., Vijayan, V., {et~al.} 2023, The Astrophysical Journal Letters, 954, L41, \dodoi{10.3847/2041-8213/acf29a}

\bibitem[{{Soares-Santos} {et~al.}(2017){Soares-Santos}, {Holz}, {Annis}, {Chornock}, {Herner}, {Berger}, {Brout}, {Chen}, {Kessler}, {Sako}, {Allam}, {Tucker}, {Butler}, {Palmese}, {Doctor}, {Diehl}, {Frieman}, {Yanny}, {Lin}, {Scolnic}, {Cowperthwaite}, {Neilsen}, {Marriner}, {Kuropatkin}, {Hartley}, {Paz-Chinch{\'o}n}, {Alexander}, {Balbinot}, {Blanchard}, {Brown}, {Carlin}, {Conselice}, {Cook}, {Drlica-Wagner}, {Drout}, {Durret}, {Eftekhari}, {Farr}, {Finley}, {Foley}, {Fong}, {Fryer}, {Garc{\'\i}a-Bellido}, {Gill}, {Gruendl}, {Hanna}, {Kasen}, {Li}, {Lopes}, {Louren{\c{c}}o}, {Margutti}, {Marshall}, {Matheson}, {Medina}, {Metzger}, {Mu{\~n}oz}, {Muir}, {Nicholl}, {Quataert}, {Rest}, {Sauseda}, {Schlegel}, {Secco}, {Sobreira}, {Stebbins}, {Villar}, {Vivas}, {Walker}, {Wester}, {Williams}, {Zenteno}, {Zhang}, {Abbott}, {Abdalla}, {Banerji}, {Bechtol}, {Benoit-L{\'e}vy}, {Bertin}, {Brooks}, {Buckley-Geer}, {Burke}, {Carnero Rosell}, {Carrasco Kind}, {Carretero}, {Castander}, {Crocce}, {Cunha}, {D'Andrea},
  {da Costa}, {Davis}, {Desai}, {Dietrich}, {Doel}, {Eifler}, {Fernandez}, {Flaugher}, {Fosalba}, {Gaztanaga}, {Gerdes}, {Giannantonio}, {Goldstein}, {Gruen}, {Gschwend}, {Gutierrez}, {Honscheid}, {Jain}, {James}, {Jeltema}, {Johnson}, {Johnson}, {Kent}, {Krause}, {Kron}, {Kuehn}, {Kuhlmann}, {Lahav}, {Lima}, {Maia}, {March}, {McMahon}, {Menanteau}, {Miquel}, {Mohr}, {Nichol}, {Nord}, {Ogando}, {Petravick}, {Plazas}, {Romer}, {Roodman}, {Rykoff}, {Sanchez}, {Scarpine}, {Schubnell}, {Sevilla-Noarbe}, {Smith}, {Smith}, {Suchyta}, {Swanson}, {Tarle}, {Thomas}, {Thomas}, {Troxel}, {Vikram}, {Wechsler}, {Weller}, {Dark Energy Survey}, \& {Dark Energy Camera GW-EM Collaboration}}]{Soaressantos2017ApJ...848L..16S}
{Soares-Santos}, M., {Holz}, D.~E., {Annis}, J., {et~al.} 2017, The Astrophysical Journal Letters, 848, L16, \dodoi{10.3847/2041-8213/aa9059}

\bibitem[{{Soares-Santos} {et~al.}(2019){Soares-Santos}, {Palmese}, {Hartley}, {Annis}, {Garcia-Bellido}, {Lahav}, {Doctor}, {Fishbach}, {Holz}, {Lin}, \& et~al.}]{darksiren1}
{Soares-Santos}, M., {Palmese}, A., {Hartley}, W., {et~al.} 2019, apjl, 876, L7, \dodoi{10.3847/2041-8213/ab14f1}

\bibitem[{Talts {et~al.}(2020)Talts, Betancourt, Simpson, Vehtari, \& Gelman}]{talts2020validating}
Talts, S., Betancourt, M., Simpson, D., Vehtari, A., \& Gelman, A. 2020, Validating Bayesian Inference Algorithms with Simulation-Based Calibration.
\newblock \doarXiv{1804.06788}

\bibitem[{Tejero-Cantero {et~al.}(2020)Tejero-Cantero, Boelts, Deistler, Lueckmann, Durkan, Gonçalves, Greenberg, \& Macke}]{SBItejero-cantero2020sbi}
Tejero-Cantero, A., Boelts, J., Deistler, M., {et~al.} 2020, Journal of Open Source Software, 5, 2505, \dodoi{10.21105/joss.02505}

\bibitem[{Yang {et~al.}(2024)Yang, Troja, O’Connor, Fryer, Im, Durbak, Paek, Ricci, Bom, Gillanders, {et~al.}}]{yang2024lanthanide}
Yang, Y.-H., Troja, E., O’Connor, B., {et~al.} 2024, Nature, 626, 742

\end{thebibliography}
\bibliographystyle{aasjournal}



\end{document}